\input hyperbasics
\catcode`\@=11
\def\unredoffs{\voffset=13mm \hoffset=6.5truemm} 
\def\redoffs{\voffset=-12.truemm\hoffset=-3truemm} 
\def\speclscape{}
%
\newbox\leftpage \newdimen\fullhsize \newdimen\hstitle \newdimen\hsbody
\newdimen\hdim
\hfuzz=1pt
\ifx\hyperdef\UNd@FiNeD\def\hyperdef#1#2#3#4{#4}\def\hyperref#1#2#3#4{#4}\fi
\def\newans{y }
\def\answb{y }
\ifx\answb\newans\message{(This uses normal fonts.)}%
%
\def\bigans{b }
\def\answ{b }
\ifx\answ\bigans\message{(Format simple colonne 12pts.}
\magnification=1200 \unredoffs\hsize=147truemm\vsize=219truemm
\hsbody=\hsize \hstitle=\hsize 
\else\message{(Format double colonne, 10pts.} \let\l@r=L
\magnification=1000 \vsize=182.5truemm
\redoffs%
\hstitle=122.5truemm\hsbody=122.5truemm\fullhsize=258truemm\hsize=\hsbody 
\output={
  \almostshipout{\leftline{\vbox{\makeheadline\pagebody\makefootline}}}
\advancepageno%
}
\def\almostshipout#1{\if L\l@r \count1=1 \message{[\the\count0.\the\count1]}
      \global\setbox\leftpage=#1 \global\let\l@r=R
 \else \count1=2
  \shipout\vbox{\speclscape{\hsize\fullhsize}
      \hbox to\fullhsize{\box\leftpage\hfil#1}}  \global\let\l@r=L\fi}
\fi

\def\sla#1{\mkern-1.5mu\raise0.4pt\hbox{$\not$}\mkern1.2mu #1\mkern 0.7mu}
\def\Dbar{\mkern-1.5mu\raise0.4pt\hbox{$\not$}\mkern-.1mu {\rm D}\mkern.1mu}
\def\Abar{\mkern1.mu\raise0.4pt\hbox{$\not$}\mkern-1.3mu A\mkern.1mu}
\def\Bbar{\mkern-0.mu\raise0.4pt\hbox{$\not$}\mkern-.3mu B\mkern 0.6mu}
\newskip\tableskipamount \tableskipamount=8pt plus 3pt minus 3pt
\def\tableskip{\vskip\tableskipamount}

\newdimen\chapskip

\font\ssbx=cmssbx10  

\font\caprm=cmr9
\font\capit=cmti9
\font\capbf=cmbx9
\font\capsl=cmsl9
\font\capmi=cmmi9
\font\capex=cmex9
\font\capsy=cmsy9
\chapskip=17.5mm
\def\makeheadline{\vbox to 0pt{\vskip-22.5pt
\line{\vbox to8.5pt{}\the\headline}\vss}\nointerlineskip}
\font\tenbi=cmmib10 
\font\ninebi=cmmib9
\font\sevenbi=cmmib7 
\font\fivebi=cmmib5
\textfont4=\tenbi
\scriptfont4=\sevenbi
\scriptscriptfont4=\fivebi
\font\headrm=cmr10

\font\sixrm=cmr6
\font\fiverm=cmr5
\font\sixmi=cmmi6
\font\fivemi=cmmi5
\font\sixsy=cmsy6
\font\fivesy=cmsy5
\font\sixbf=cmbx6
\font\fivebf=cmbx5
\skewchar\capmi='177 \skewchar\sixmi='177 \skewchar\fivemi='177
\skewchar\capsy='60 \skewchar\sixsy='60 \skewchar\fivesy='60

\def\elevenpoint{
\textfont0=\caprm \scriptfont0=\sixrm \scriptscriptfont0=\fiverm
\def\rm{\fam0\caprm}
\textfont1=\capmi \scriptfont1=\sixmi \scriptscriptfont1=\fivemi
\textfont2=\capsy \scriptfont2=\sixsy \scriptscriptfont2=\fivesy
\textfont3=\capex \scriptfont3=\capex \scriptscriptfont3=\capex
\textfont\itfam=\capit \def\it{\fam\itfam\capit} 
\textfont\slfam=\capsl  \def\sl{\fam\slfam\capsl} 
\textfont\bffam=\capbf \scriptfont\bffam=\sixbf
\scriptscriptfont\bffam=\fivebf
\def\bf{\fam\bffam\capbf} 
\textfont4=\ninebi \scriptfont4=\sevenbi \scriptscriptfont4=\fivebi
\abovedisplayskip=11pt plus 3pt minus 8pt
\belowdisplayskip=\abovedisplayskip
\smallskipamount=2.7pt plus 1pt minus 1pt
\medskipamount=5.4pt plus 2pt minus 2pt
\bigskipamount=10.8pt plus 3.6pt minus 3.6pt
\normalbaselineskip=11pt
\setbox\strutbox=\hbox{\vrule height7.8pt depth3.2pt width0pt}
\normalbaselines \rm}

%
%

\catcode`\@=11

\font\tenmsa=msam10
\font\sevenmsa=msam7
\font\fivemsa=msam5
\font\tenmsb=msbm10
\font\sevenmsb=msbm7
\font\fivemsb=msbm5
\newfam\msafam
\newfam\msbfam
\textfont\msafam=\tenmsa  \scriptfont\msafam=\sevenmsa
  \scriptscriptfont\msafam=\fivemsa
\textfont\msbfam=\tenmsb  \scriptfont\msbfam=\sevenmsb
  \scriptscriptfont\msbfam=\fivemsb

\def\hexnumber@#1{\ifcase#1 0\or1\or2\or3\or4\or5\or6\or7\or8\or9\or
	A\or B\or C\or D\or E\or F\fi }

\font\teneuf=eufm10
\font\seveneuf=eufm7
\font\fiveeuf=eufm5
\newfam\euffam
\textfont\euffam=\teneuf
\scriptfont\euffam=\seveneuf
\scriptscriptfont\euffam=\fiveeuf
\def\frak{\ifmmode\let\next\frak@\else
 \def\next{\Err@{Use \string\frak\space only in math mode}}\fi\next}
\def\goth{\ifmmode\let\next\frak@\else
 \def\next{\Err@{Use \string\goth\space only in math mode}}\fi\next}
\def\frak@#1{{\frak@@{#1}}}
\def\frak@@#1{\fam\euffam#1}

\edef\msa@{\hexnumber@\msafam}
\edef\msb@{\hexnumber@\msbfam}

\mathchardef\boxdot="2\msa@00
\mathchardef\boxplus="2\msa@01
\mathchardef\boxtimes="2\msa@02
\mathchardef\square="0\msa@03
\mathchardef\blacksquare="0\msa@04
\mathchardef\centerdot="2\msa@05
\mathchardef\lozenge="0\msa@06
\mathchardef\blacklozenge="0\msa@07
\mathchardef\circlearrowright="3\msa@08
\mathchardef\circlearrowleft="3\msa@09
\mathchardef\rightleftharpoons="3\msa@0A
\mathchardef\leftrightharpoons="3\msa@0B
\mathchardef\boxminus="2\msa@0C
\mathchardef\Vdash="3\msa@0D
\mathchardef\Vvdash="3\msa@0E
\mathchardef\vDash="3\msa@0F
\mathchardef\twoheadrightarrow="3\msa@10
\mathchardef\twoheadleftarrow="3\msa@11
\mathchardef\leftleftarrows="3\msa@12
\mathchardef\rightrightarrows="3\msa@13
\mathchardef\upuparrows="3\msa@14
\mathchardef\downdownarrows="3\msa@15
\mathchardef\upharpoonright="3\msa@16

\mathchardef\downharpoonright="3\msa@17
\mathchardef\upharpoonleft="3\msa@18
\mathchardef\downharpoonleft="3\msa@19
\mathchardef\rightarrowtail="3\msa@1A
\mathchardef\leftarrowtail="3\msa@1B
\mathchardef\leftrightarrows="3\msa@1C
\mathchardef\rightleftarrows="3\msa@1D
\mathchardef\Lsh="3\msa@1E
\mathchardef\Rsh="3\msa@1F
\mathchardef\rightsquigarrow="3\msa@20
\mathchardef\leftrightsquigarrow="3\msa@21
\mathchardef\looparrowleft="3\msa@22
\mathchardef\looparrowright="3\msa@23
\mathchardef\circeq="3\msa@24
\mathchardef\succsim="3\msa@25
\mathchardef\gtrsim="3\msa@26
\mathchardef\gtrapprox="3\msa@27
\mathchardef\multimap="3\msa@28
\mathchardef\therefore="3\msa@29
\mathchardef\because="3\msa@2A
\mathchardef\doteqdot="3\msa@2B

\mathchardef\triangleq="3\msa@2C
\mathchardef\precsim="3\msa@2D
\mathchardef\lesssim="3\msa@2E
\mathchardef\lessapprox="3\msa@2F
\mathchardef\eqslantless="3\msa@30
\mathchardef\eqslantgtr="3\msa@31
\mathchardef\curlyeqprec="3\msa@32
\mathchardef\curlyeqsucc="3\msa@33
\mathchardef\preccurlyeq="3\msa@34
\mathchardef\leqq="3\msa@35
\mathchardef\leqslant="3\msa@36
\mathchardef\lessgtr="3\msa@37
\mathchardef\backprime="0\msa@38
\mathchardef\risingdotseq="3\msa@3A
\mathchardef\fallingdotseq="3\msa@3B
\mathchardef\succcurlyeq="3\msa@3C
\mathchardef\geqq="3\msa@3D
\mathchardef\geqslant="3\msa@3E
\mathchardef\gtrless="3\msa@3F
\mathchardef\sqsubset="3\msa@40
\mathchardef\sqsupset="3\msa@41
\mathchardef\vartriangleright="3\msa@42
\mathchardef\vartriangleleft="3\msa@43
\mathchardef\trianglerighteq="3\msa@44
\mathchardef\trianglelefteq="3\msa@45
\mathchardef\bigstar="0\msa@46
\mathchardef\between="3\msa@47
\mathchardef\blacktriangledown="0\msa@48
\mathchardef\blacktriangleright="3\msa@49
\mathchardef\blacktriangleleft="3\msa@4A
\mathchardef\vartriangle="0\msa@4D
\mathchardef\blacktriangle="0\msa@4E
\mathchardef\triangledown="0\msa@4F
\mathchardef\eqcirc="3\msa@50
\mathchardef\lesseqgtr="3\msa@51
\mathchardef\gtreqless="3\msa@52
\mathchardef\lesseqqgtr="3\msa@53
\mathchardef\gtreqqless="3\msa@54
\mathchardef\Rrightarrow="3\msa@56
\mathchardef\Lleftarrow="3\msa@57
\mathchardef\veebar="2\msa@59
\mathchardef\barwedge="2\msa@5A
\mathchardef\doublebarwedge="2\msa@5B
\mathchardef\angle="0\msa@5C
\mathchardef\measuredangle="0\msa@5D
\mathchardef\sphericalangle="0\msa@5E
\mathchardef\varpropto="3\msa@5F
\mathchardef\smallsmile="3\msa@60
\mathchardef\smallfrown="3\msa@61
\mathchardef\Subset="3\msa@62
\mathchardef\Supset="3\msa@63
\mathchardef\Cup="2\msa@64

\mathchardef\Cap="2\msa@65

\mathchardef\curlywedge="2\msa@66
\mathchardef\curlyvee="2\msa@67
\mathchardef\leftthreetimes="2\msa@68
\mathchardef\rightthreetimes="2\msa@69
\mathchardef\subseteqq="3\msa@6A
\mathchardef\supseteqq="3\msa@6B
\mathchardef\bumpeq="3\msa@6C
\mathchardef\Bumpeq="3\msa@6D
\mathchardef\lll="3\msa@6E

\mathchardef\ggg="3\msa@6F

\mathchardef\circledS="0\msa@73
\mathchardef\pitchfork="3\msa@74
\mathchardef\dotplus="2\msa@75
\mathchardef\backsim="3\msa@76
\mathchardef\backsimeq="3\msa@77
\mathchardef\complement="0\msa@7B
\mathchardef\intercal="2\msa@7C
\mathchardef\circledcirc="2\msa@7D
\mathchardef\circledast="2\msa@7E
\mathchardef\circleddash="2\msa@7F
\def\ulcorner{\delimiter"4\msa@70\msa@70 }
\def\urcorner{\delimiter"5\msa@71\msa@71 }
\def\llcorner{\delimiter"4\msa@78\msa@78 }
\def\lrcorner{\delimiter"5\msa@79\msa@79 }
\def\yen{\mathhexbox\msa@55 }
\def\checkmark{\mathhexbox\msa@58 }
\def\circledR{\mathhexbox\msa@72 }
\def\maltese{\mathhexbox\msa@7A }
\mathchardef\lvertneqq="3\msb@00
\mathchardef\gvertneqq="3\msb@01
\mathchardef\nleq="3\msb@02
\mathchardef\ngeq="3\msb@03
\mathchardef\nless="3\msb@04
\mathchardef\ngtr="3\msb@05
\mathchardef\nprec="3\msb@06
\mathchardef\nsucc="3\msb@07
\mathchardef\lneqq="3\msb@08
\mathchardef\gneqq="3\msb@09
\mathchardef\nleqslant="3\msb@0A
\mathchardef\ngeqslant="3\msb@0B
\mathchardef\lneq="3\msb@0C
\mathchardef\gneq="3\msb@0D
\mathchardef\npreceq="3\msb@0E
\mathchardef\nsucceq="3\msb@0F
\mathchardef\precnsim="3\msb@10
\mathchardef\succnsim="3\msb@11
\mathchardef\lnsim="3\msb@12
\mathchardef\gnsim="3\msb@13
\mathchardef\nleqq="3\msb@14
\mathchardef\ngeqq="3\msb@15
\mathchardef\precneqq="3\msb@16
\mathchardef\succneqq="3\msb@17
\mathchardef\precnapprox="3\msb@18
\mathchardef\succnapprox="3\msb@19
\mathchardef\lnapprox="3\msb@1A
\mathchardef\gnapprox="3\msb@1B
\mathchardef\nsim="3\msb@1C
\mathchardef\ncong="3\msb@1D

\mathchardef\varsubsetneq="3\msb@20
\mathchardef\varsupsetneq="3\msb@21
\mathchardef\nsubseteqq="3\msb@22
\mathchardef\nsupseteqq="3\msb@23
\mathchardef\subsetneqq="3\msb@24
\mathchardef\supsetneqq="3\msb@25
\mathchardef\varsubsetneqq="3\msb@26
\mathchardef\varsupsetneqq="3\msb@27
\mathchardef\subsetneq="3\msb@28
\mathchardef\supsetneq="3\msb@29
\mathchardef\nsubseteq="3\msb@2A
\mathchardef\nsupseteq="3\msb@2B
\mathchardef\nparallel="3\msb@2C
\mathchardef\nmid="3\msb@2D
\mathchardef\nshortmid="3\msb@2E
\mathchardef\nshortparallel="3\msb@2F
\mathchardef\nvdash="3\msb@30
\mathchardef\nVdash="3\msb@31
\mathchardef\nvDash="3\msb@32
\mathchardef\nVDash="3\msb@33
\mathchardef\ntrianglerighteq="3\msb@34
\mathchardef\ntrianglelefteq="3\msb@35
\mathchardef\ntriangleleft="3\msb@36
\mathchardef\ntriangleright="3\msb@37
\mathchardef\nleftarrow="3\msb@38
\mathchardef\nrightarrow="3\msb@39
\mathchardef\nLeftarrow="3\msb@3A
\mathchardef\nRightarrow="3\msb@3B
\mathchardef\nLeftrightarrow="3\msb@3C
\mathchardef\nleftrightarrow="3\msb@3D
\mathchardef\divideontimes="2\msb@3E
\mathchardef\varnothing="0\msb@3F
\mathchardef\nexists="0\msb@40
\mathchardef\mho="0\msb@66
\mathchardef\eth="0\msb@67
\mathchardef\eqsim="3\msb@68
\mathchardef\beth="0\msb@69
\mathchardef\gimel="0\msb@6A
\mathchardef\daleth="0\msb@6B
\mathchardef\lessdot="3\msb@6C
\mathchardef\gtrdot="3\msb@6D
\mathchardef\ltimes="2\msb@6E
\mathchardef\rtimes="2\msb@6F
\mathchardef\shortmid="3\msb@70
\mathchardef\shortparallel="3\msb@71
\mathchardef\smallsetminus="2\msb@72
\mathchardef\thicksim="3\msb@73
\mathchardef\thickapprox="3\msb@74
\mathchardef\approxeq="3\msb@75
\mathchardef\succapprox="3\msb@76
\mathchardef\precapprox="3\msb@77
\mathchardef\curvearrowleft="3\msb@78
\mathchardef\curvearrowright="3\msb@79
\mathchardef\digamma="0\msb@7A
\mathchardef\varkappa="0\msb@7B
\mathchardef\hslash="0\msb@7D
\mathchardef\hbar="0\msb@7E
\mathchardef\backepsilon="3\msb@7F
\def\Bbb{\ifmmode\let\next\Bbb@\else
 \def\next{\errmessage{Use \string\Bbb\space only in math mode}}\fi\next}
\def\Bbb@#1{{\Bbb@@{#1}}}
\def\Bbb@@#1{\fam\msbfam#1}

\catcode`\@=12

\def\sla#1{\mkern-1.5mu\raise0.4pt\hbox{$\not$}\mkern1.2mu #1\mkern 0.7mu}
\def\Dbar{\mkern-1.5mu\raise0.4pt\hbox{$\not$}\mkern-.1mu {\rm D}\mkern.1mu}
\def\Abar{\mkern1.mu\raise0.4pt\hbox{$\not$}\mkern-1.3mu A\mkern.1mu}
\nopagenumbers
\headline={\ifnum\pageno=1\hfill\else\draftdate\hfil{\headrm\folio}%
\hfil\fi}	 
\else\message{(This uses pseudo 12pts fonts.}
\hoffset=8mm
\voffset=16mm
\input lfont12 

\def\sla#1{\mkern-1.5mu\raise0.5pt\hbox{$\not$}\mkern1.2mu #1\mkern 0.7mu}
\def\Dbar{\mkern-1.5mu\raise0.5pt\hbox{$\not$}\mkern-.1mu {\rm D}\mkern.1mu}
\def\Abar{\mkern1.mu\raise0.5pt\hbox{$\not$}\mkern-1.3mu A\mkern.1mu}
\fi
\def\fileth{\noalign{\hrule}}

\newcount\yearltd\yearltd=\year\advance\yearltd by -1900
\newif\ifdraftmode
\draftmodefalse
\def\draft{\draftmodetrue{\count255=\time\divide\count255 by 60
\xdef\hourmin{\number\count255} 
  \multiply\count255 by-60\advance\count255 by\time
  \xdef\hourmin{\hourmin:\ifnum\count255<10 0\fi\the\count255}}}
\def\draftdate{\ifdraftmode{\headrm\quad (\jobname,\ le
\number\day/\number\month/\number\yearltd\ \ \hourmin)}\else{}\fi} 
\newif\iffrancmode
\francmodefalse
\def\e{\mathop{\rm e}\nolimits}

\def\d{{\rm d}}
\def\ud{{\textstyle{1\over 2}}}

\def\del{\partial}

\chardef\sigmat=27

\def\frac#1#2{{\textstyle{#1\over#2}}}

\def\leaderfill{\leaders\hbox to 1em{\hss.\hss}\hfill}
\catcode`\@=11
\def\deqalignno#1{\displ@y\tabskip\centering \halign to
\displaywidth{\hfil$\displaystyle{##}$\tabskip0pt&$\displaystyle{{}##}$
\hfil\tabskip0pt &\quad
\hfil$\displaystyle{##}$\tabskip0pt&$\displaystyle{{}##}$ 
\hfil\tabskip\centering& \llap{$##$}\tabskip0pt \crcr #1 \crcr}}
\def\deqalign#1{\null\,\vcenter{\openup\jot\m@th\ialign{
\strut\hfil$\displaystyle{##}$&$\displaystyle{{}##}$\hfil
&&\quad\strut\hfil$\displaystyle{##}$&$\displaystyle{{}##}$
\hfil\crcr#1\crcr}}\,}
\def\xlabel#1{\expandafter\xl@bel#1}\def\xl@bel#1{#1}
\def\label#1{\l@bel #1{\hbox{}}}
\def\l@bel#1{\ifx\UNd@FiNeD#1\message{label \string#1 is undefined.}%
\xdef#1{?.? }\fi{\let\hyperref=\relax\xdef\next{#1}}%
\ifx\next\em@rk\def\next{}%
\else\def\next{#1}\fi\next}
\def\DefWarn#1{\ifx\UNd@FiNeD#1\else
\immediate\write16{*** WARNING: the label \string#1 is already defined%
***}\fi}%
\newread\ch@ckfile
\def\cinput#1{\def\filen@me{#1 }
\immediate\openin\ch@ckfile=\filen@me
\ifeof\ch@ckfile\message{<< (\filen@me\ DOES NOT EXIST in this pass)>>}\else%
\closein \ch@ckfile\input\filen@me\fi}
\ifx\UNd@FiNeD\prefix\def\prefix{}\fi 
\newread\ch@ckfile
\immediate\openin\ch@ckfile=\jobname.def
\ifeof\ch@ckfile\message{<< (\jobname.def DOES NOT EXIST in this pass) >>}
\else
\def\DefWarn#1{}%
\closein \ch@ckfile%
\input\jobname.def\fi
\def\listcontent{
\immediate\openin\ch@ckfile=\jobname.tab 
\ifeof\ch@ckfile\message{no file \jobname.tab, no table of contents this
pass}%
\else\closein\ch@ckfile\centerline{\bf\iffrancmode Table des
mati\`eres \else Contents\fi}\nobreak\medskip%
{\baselineskip=12pt\parskip=0pt\catcode`\@=11\input\jobname.tab
\catcode`\@=12\bigbreak\bigskip}\fi}
\newcount\nosection
\newcount\nosubsection
\newcount\neqno
\newcount\notenumber
\newcount\nofigure
\newcount\notable
\newif\ifappmode
\def\equation{\jobname.equ}
\newwrite\equa

\newdimen\hulp
\def\maketitle#1{
\edef\oneliner##1{\centerline{##1}}
\edef\twoliner##1{\vbox{\parindent=0pt\leftskip=0pt plus 1fill\rightskip=0pt
plus 1fill 
                     \parfillskip=0pt\relax##1}} 
\setbox0=\vbox{#1}\hulp=0.5\hsize
                 \ifdim\wd0<\hulp\oneliner{#1}\else
                 \twoliner{#1}\fi}
\def\preprint#1{\ifdraftmode\gdef\prepname{\jobname/#1}\else%
\gdef\prepname{#1}\fi\hfill{
\expandafter{\prepname}}\vskip20mm} 
\def\title#1\par{\gdef\titlename{#1}
\maketitle{\ssbx\uppercase\expandafter{\titlename}}
\vskip20truemm
\nosection=0
\neqno=0
\notenumber=0
\nofigure=0
\notable=0
\def\prefix{}
\appmodefalse
\mark{\the\nosection}
\message{#1}
\immediate\openout\equa=\equation
}
\def\abstract{\vskip8mm\iffrancmode\centerline{R\'ESUM\'E}\else%
\centerline{ABSTRACT}\fi \smallskip \begingroup\narrower
\elevenpoint\baselineskip10pt} 
\def\endabstract{\par\endgroup \bigskip}
\def\section#1\par{\vskip0pt plus.1\vsize\penalty-100\vskip0pt plus-.1
\vsize\bigskip\vskip\parskip
\ifnum\nosection=0\ifappmode\relax\else\writetoc
\fi\fi
\advance\nosection by 1\global\nosubsection=0\global\neqno=0
\vbox{\noindent\bf{\hyperdef\hypernoname{section}{\prefix\the\nosection}%
{\prefix\the\nosection}\ #1}}
\writetoca{{\string\hyperref{}{section}{\prefix\the\nosection}%
{\prefix\the\nosection}} {#1}}
\message{\the\nosection\ #1}
\mark{\the\nosection}\bigskip\noindent
}

\def\appendix#1#2\par{\bigbreak\nosection=0
\appmodetrue
\notenumber=0
\neqno=0
\def\prefix{A}
\mark{\the\nosection}
\message{APPENDICES}
{\leftline{APPENDICES} \hyperdef\hypernoname{appendix}{\prefix}{ 
\leftline{\uppercase\expandafter{#1}}
\leftline{\uppercase\expandafter{#2}}}}
\bigskip\noindent\nonfrenchspacing
\writetoca{\string\hyperref{}{appendix}{\prefix}{Appendices}.\ #1.\ #2}%
}
\def\subsection#1\par {\vskip0pt plus.05\vsize\penalty-100\vskip0pt
plus-.05\vsize\bigskip\vskip\parskip\advance\nosubsection by 1
\vbox{\noindent\it{\hyperdef\hypernoname{subsection}{\prefix\the\nosection.%
\the\nosubsection}{\prefix\the\nosection.\the\nosubsection\ #1}}}%
\smallskip\noindent 
\writetoca{{\string\hyperref{}{subsection}{\prefix\the\nosection.%
\the\nosubsection}{\prefix\the\nosection.\the\nosubsection}} {#1}}
\message{\the\nosection.\the\nosubsection\ #1}
} 
\def\note #1{\advance\notenumber by 1
\footnote{$^{\the\notenumber}$}{\sevenrm #1}} 

\parindent=1em 
\newinsert\margin
\dimen\margin=\maxdimen
\count\margin=0 \skip\margin=0pt
\def\sslbl#1{\DefWarn#1%
\ifdraftmode{\hfill\escapechar-1{\rlap{\hskip-1mm%
\sevenrm\string#1}}}\fi%
\ifnum\nosection=0\if\prefix{}\xdef#1{}%
\edef\ewrite{\write\equa{{\string#1}}%
\write\equa{}}\ewrite%
\else
\xdef#1{\noexpand\hyperref{}{appendix}{\prefix}{\prefix}}%
\edef\ewrite{\write\equa{{\string#1},\prefix}%
\write\equa{}}\ewrite%
\writedef{#1\leftbracket#1}
\fi
\else%
\ifnum\nosubsection=0%
\xdef#1{\noexpand\hyperref{}{section}{\prefix\the\nosection}%
{\prefix\the\nosection}}%
\edef\ewrite{\write\equa{{\string#1},\prefix\the\nosection}%
\write\equa{}}\ewrite%
\writedef{#1\leftbracket#1}
\else%
\xdef#1{\noexpand\hyperref{}{subsection}{\prefix\the\nosection.%
\the\nosubsection}{\prefix\the\nosection.\the\nosubsection}}%
\writedef{#1\leftbracket#1}
\edef\ewrite{\write\equa{{\string#1},\prefix\the\nosection%
.\the\nosubsection}\write\equa{}}\ewrite\fi\fi}%

\newwrite\tfile \def\writetoca#1{}
\def\writetoc{\immediate\openout\tfile=\jobname.tab
\def\writetoca##1{{\edef\next{\write\tfile{\noindent ##1 \string\leaderfill%
\noexpand\number\pageno\par}}\next}}}

%
\def\nolabels{\def\wrlabeL##1{}\def\eqlabeL##1{}\def\reflabeL##1{}}
\def\writelabels{\def\wrlabeL##1{\leavevmode\vadjust{\rlap{\smash%
{\line{{\escapechar=` \hfill\rlap{\sevenrm\hskip.03in\string##1}}}}}}}%
\def\eqlabeL##1{{\escapechar-1\rlap{\sevenrm\hskip.05in\string##1}}}%
\def\reflabeL##1{\noexpand\llap{\noexpand\sevenrm\string\string\string##1}}}
\nolabels

\global\newcount\refno \global\refno=1
\newwrite\rfile
\def\ref{[\hyperref{}{reference}{\the\refno}{\the\refno}]\nref}
\def\nref#1{\DefWarn#1%
\xdef#1{[\noexpand\hyperref{}{reference}{\the\refno}{\the\refno}]}%
\writedef{#1\leftbracket#1}%
\ifnum\refno=1\immediate\openout\rfile=\jobname.ref\fi
\chardef\wfile=\rfile\immediate\write\rfile{\noexpand\item{[\noexpand\hyperdef%
\noexpand\hypernoname{reference}{\the\refno}{\the\refno}]\ }%
\reflabeL{#1\hskip.31in}\pctsign}\global\advance\refno by1\findarg}
\def\findarg#1#{\begingroup\obeylines\newlinechar=`\^^M\pass@rg}
{\obeylines\gdef\pass@rg#1{\writ@line\relax #1^^M\hbox{}^^M}%
\gdef\writ@line#1^^M{\expandafter\toks0\expandafter{\striprel@x #1}%
\edef\next{\the\toks0}\ifx\next\em@rk\let\next=\endgroup\else\ifx\next\empty%
\else\immediate\write\wfile{\the\toks0}\fi\let\next=\writ@line\fi\next\relax}}
\def\striprel@x#1{} \def\em@rk{\hbox{}}
\def\lref{\begingroup\obeylines\lr@f}
\def\lr@f#1#2{\DefWarn#1\gdef#1{\let#1=\UNd@FiNeD\ref#1{#2}}\endgroup\unskip}
\def\semi{;\hfil\break}
\def\addref#1{\immediate\write\rfile{\noexpand\item{}#1}} 
\def\listrefs{{}\vfill\supereject\immediate\closeout\rfile\writestoppt
\baselineskip=14pt\centerline{{\bf\iffrancmode R\'eferences\else References%
\fi}}
\bigskip{\parindent=20pt%
\frenchspacing\escapechar=` \input \jobname.ref\vfill\eject}\nonfrenchspacing}
\def\startrefs#1{\immediate\openout\rfile=\jobname.ref\refno=#1}
\def\xref{\expandafter\xr@f}\def\xr@f[#1]{#1}
\def\refs#1{\count255=1[\r@fs #1{\hbox{}}]}
\def\r@fs#1{\ifx\UNd@FiNeD#1\message{reflabel \string#1 is undefined.}%
\nref#1{need to supply reference \string#1.}\fi%
\vphantom{\hphantom{#1}}{\let\hyperref=\relax\xdef\next{#1}}%
\ifx\next\em@rk\def\next{}%
\else\ifx\next#1\ifodd\count255\relax\xref#1\count255=0\fi%
\else#1\count255=1\fi\let\next=\r@fs\fi\next}
%
\newwrite\lfile
{\escapechar-1\xdef\pctsign{\string\%}\xdef\leftbracket{\string\{}
\xdef\rightbracket{\string\}}\xdef\numbersign{\string\#}}
\def\writedefs{\immediate\openout\lfile=\jobname.def \def\writedef##1{%
{\let\hyperref=\relax\let\hyperdef=\relax\let\hypernoname=\relax
 \immediate\write\lfile{\string\def\string##1\rightbracket}}}}%
\def\writestop{\def\writestoppt{\immediate\write\lfile{\string\pageno%
\the\pageno\string\startrefs\leftbracket\the\refno\rightbracket%
\string\def\string\secsym\leftbracket\secsym\rightbracket%
\string\secno\the\secno\string\meqno\the\meqno}\immediate\closeout\lfile}}
\def\writestoppt{}\def\writedef#1{}
\writedefs
\def\biblio\par{\vskip0pt plus.1\vsize\penalty-100\vskip0pt plus-.1
\vsize\bigskip\vskip\parskip
\message{Bibliographie}
{\leftline{\bf \hyperdef\hypernoname{biblio}{bib}{Bibliographical Notes}}}
\nobreak\medskip\noindent\frenchspacing
\writetoca{\string\hyperref{}{biblio}{bib}{Bibliographical Notes}}}%

\def\biblionote{\iffrancmode Notes Bibliographiques\else Bibliographical Notes
\fi}
\def\beginbib\par{\vskip0pt plus.1\vsize\penalty-100\vskip0pt plus-.1
\vsize\bigskip\vskip\parskip
\message{Bibliographie}
{\leftline{\bf \hyperdef\hypernoname{biblio}{\the\nosection}%
{\biblionote}}}
\nobreak\medskip\noindent\frenchspacing
\writetoca{\string\hyperref{}{biblio}{\the\nosection}%
{\biblionote}}}%

\def\Exercises{\iffrancmode Exercices\else Exercises
\fi}
\def\exerc\par{\vskip0pt plus.1\vsize\penalty-100\vskip0pt plus-.1
\vsize\bigskip\vskip\parskip
\message{Exercises}
{\leftline{\bf \hyperdef\hypernoname{exercise}{\the\nosection}{\Exercises}}}
\nobreak\medskip\noindent\frenchspacing
\writetoca{\string\hyperref{}{exercise}{\the\nosection}{\Exercises}}
}
\def\eqnn{\global\advance\neqno by 1 \ifinner\relax\else%
\eqno\fi(\prefix\the\nosection.\the\neqno)}
%
\def\eqnd#1{\DefWarn#1%
\global\advance\neqno by 1 
{\xdef#1{($\noexpand\hyperref{}{equation}{\prefix\the\nosection.\the\neqno}%
{\prefix\the\nosection.\the\neqno}$)}}
\ifinner\relax\else\eqno\fi(\hyperdef\hypernoname{equation}{\prefix\the%
\nosection.\the\neqno}{\prefix\the\nosection.\the\neqno})
\writedef{#1\leftbracket#1}
\ifdraftmode{\escapechar-1{\rlap{\hskip.2mm\sevenrm\string#1}}}\fi
\edef\ewrite{\write\equa{{\string#1},(\prefix\the\nosection.\the\neqno)
{\noexpand\number\pageno}}\write\equa{}}\ewrite}
%
\def\checkm@de#1#2{\ifmmode{\def\f@rst##1{##1}\hyperdef\hypernoname{equation}%
{#1}{#2}}\else\hyperref{}{equation}{#1}{#2}\fi}
\def\f@rst#1{\c@t#1a\em@ark}\def\c@t#1#2\em@ark{#1}
\def\eqna#1{\DefWarn#1%
\global\advance\neqno by1\ifdraftmode{\hfill%
\escapechar-1{\rlap{\sevenrm\string#1}}}\fi%
\xdef #1##1{(\noexpand\relax\noexpand%
\checkm@de{\prefix\the\nosection.\the\neqno\noexpand\f@rst{##1}1}%
{\hbox{$\prefix\the\nosection.\the\neqno##1$}})}
\writedef{#1\numbersign1\leftbracket#1{\numbersign1}}%
} 
%

%
\def\em@rk{\hbox{}} 
\def\xeqn{\expandafter\xe@n}\def\xe@n(#1){#1}
\def\xeqna#1{\expandafter\xe@na#1}\def\xe@na\hbox#1{\xe@nap #1}
\def\xe@nap$(#1)${\hbox{$#1$}}
\def\eqns#1{(\e@ns #1{\hbox{}})}
\def\e@ns#1{\ifx\UNd@FiNeD#1\message{eqnlabel \string#1 is undefined.}%
\xdef#1{(?.?)}\fi{\let\hyperref=\relax\xdef\next{#1}}%
\ifx\next\em@rk\def\next{}%
\else\ifx\next#1\xeqn#1\else\def\n@xt{#1}\ifx\n@xt\next#1\else\xeqna#1\fi
\fi\let\next=\e@ns\fi\next}
\def\figure#1#2{\global\advance\nofigure by 1 \vglue#1%
\hyperdef\hypernoname{figure}{\the\nofigure}{}%
{\elevenpoint
\setbox1=\hbox{#2}
\ifdim\wd1=0pt\centerline{Fig.\ \the\nofigure\hskip0.5mm}%
\else\def\caption{Fig.\ \the\nofigure\quad#2\hskip0mm}
\setbox0=\hbox{\caption}
\ifdim\wd0>\hsize\noindent Fig.\ \the\nofigure\quad#2\else
                 \centerline{\caption}\fi\fi}\par}
\def\lfigure#1#2{\global\advance\nofigure by
1\vglue#1%
\hyperdef\hypernoname{figure}{\the\nofigure}{}%
\leftline{\elevenpoint\hskip10truemm  Fig.\
\the\nofigure\quad #2}} 
\def\figlbl#1{\ifdraftmode{\hfill\escapechar-1{\rlap{\hskip-1mm%
\sevenrm\string#1}}}\fi%
{\xdef#1{\noexpand\hyperref{}{figure}{\the\nofigure}%
{\the\nofigure}}}%
\edef\ewrite{\write\equa{{\string#1}\the\nofigure}%
\write\equa{}}\ewrite%
\writedef{#1\leftbracket#1}}
\def\tablbl#1{\global\advance\notable by
1\ifdraftmode{\hfill\escapechar-1{\rlap{\hskip-1mm%
\sevenrm\string#1}}}\fi%
{\xdef#1{\noexpand\hyperref{}{table}{\the\notable}%
{\the\notable}}}%
\hyperdef\hypernoname{table}{\the\notable}{}%
\edef\ewrite{\write\equa{{\string#1}\the\notable}%
\write\equa{}}\ewrite%
\writedef{#1\leftbracket#1}}

\catcode`@=12

\input epsf
\def\free{{\rm free}}
\def\bc{{\rm bc}}
\francmodefalse
\def\r{{\rm R}}
\def\F{\tilde F}
\def\g{\tilde{g}}
\preprint{SPhT/00-009}

\title{Precise determination of critical exponents and equation of state by field theory methods} 

\centerline{\caprm J.~ZINN-JUSTIN}
\medskip
{\capit\centerline{ CEA-Saclay, Service de Physique Th\'eorique*, F-91191
Gif-sur-Yvette,}\par
\centerline{ Cedex, FRANCE} \par
\centerline{E-mail: zinn@spht.saclay.cea.fr}}
 
\footnote{}{Talk given at the Conference ``Renormalization Group 2000, 
Taxco (Mexico), 11--15  Jan.~1999} 
\footnote{}{${}^{*}$Laboratoire de la Direction des
Sciences de la Mati\`ere du 
Commissariat \`a l'Energie Atomique}

\nref\rLGZJ{J.C. Le Guillou and J. Zinn-Justin, {\it Phys. Rev. Lett.} 39 (1977) 95; {\it Phys. Rev.} B21 (1980)
3976.} 
\nref\rGuiZJii{R. Guida and J. Zinn-Justin, {\it J. Phys.}  A31 (1998) 8103,
cond-mat/9803240.} 

\abstract

Renormalization group, and in particular its Quantum Field Theory
implementation has provided us with essential tools for the description of the
phase transitions and critical phenomena beyond mean field theory. We therefore review the methods, based on
renormalized $\phi^4_3$ quantum field theory and renormalization group, which
have led to a precise determination of critical exponents of the $N$-vector
model \refs{\rLGZJ,\rGuiZJii} and of the equation of state of the 3D Ising
model \ref\rGuiZJ{R.~Guida and J.~Zinn-Justin, {\it   
Nucl. Phys.} B489 [FS] (1997) 626, hep-th/9610223.}. These results are among the most precise available probing field theory in a non-perturbative regime. \par
Precise calculations first require enough terms of the perturbative expansion. However perturbation series are known to be divergent. The divergence has been characterized by relating it to instanton contributions. The information about large order behaviour of perturbation series has then allowed to develop efficient ``summation"  techniques, based on Borel transformation and conformal mapping \ref\rlob{{\it Large Order Behaviour of Perturbation Theory}, {\it Current Physics} vol.~7, J.C. Le Guillou and J. Zinn-Justin eds.,
(North-Holland, Amsterdam 1990).}. \par
We first discuss exponents  and describe our recent results \rGuiZJii. Compared to exponents, the determination of the scaling equation of state of the 3D Ising model  involves a few additional (non-trivial) technical steps, like the use of the parametric representation, and the order dependent mapping method. From the knowledge of the equation of state a number of ratio of critical amplitudes can also be derived.\par
Finally we emphasize that  few physical quantities which are
predicted by renormalization group to be {\it universal}\/ have been
determined precisely, and much work remains to be done. Considering the
steady increase in the available computer resources, many new calculations
will become feasible.  In addition to the infinite volume quantities, finite
size universal  quantities would also be of interest, to provide a more direct
contact with numerical simulations. Let us also mention dynamical observables,
a largely unexplored territory.  
\endabstract
\section Introduction: the effective $\phi\sp{4}$ field theory

Second order phase transitions are continuous phase transitions
where the correlation length diverges. Renormalization group (RG) arguments
\ref\rKWW{An early review is K.G. Wilson and J. Kogut, {\it Phys. Rep.}
12C (1974) 75.}, as well as an analysis, near dimension four, of the most IR
divergent terms appearing in the expansion around 
mean field theory \ref\rBLGZJ{E. Br\'ezin, J.C. Le Guillou and J. Zinn-Justin,
in {\it Phase Transitions and Critical Phenomena} vol.~6, C. Domb and M.S.
Green eds. (Academic Press, London 1976).}, indicate that such
transitions possess universal properties, i.e.~properties independent to a
large extent from the details of the microscopic dynamics. Moreover
all universal quantities can be calculated from renormalizable quantum field
theories which are local when the interactions are short range. For an
important class of physical systems and models  one is led to a $\phi^4$-like
euclidean field theory with $O(N)$ symmetry. Among those let us mention
statistical properties of polymers, liquid--vapour, binary mixtures,
superfluid Helium, ferromagnetic transitions...      
We explain here how critical exponents and other universal quantities have
been calculated with the help of renormalization group ideas and their
implementation within the quantum field theory context \ref\rbook{A general
reference on the subject is J.~Zinn-Justin, 1989, {\it Quantum Field Theory
and Critical Phenomena}, in particular chap.~28 of third ed., Clarendon Press
(Oxford 1989, third ed.~1996).}.\par   
\medskip
{\it The effective quantum field theory.} 
Wilson's renormalization group shows that, at least  in the neighbourhood of 
dimension four, the universal properties of statistical models which short
range interactions and $O(N)$ symmetry can generically be described by an {\it
effective}\/ euclidean field theory, with an action  $ {\cal H} \left(\phi
\right) $     
$$ {\cal H} \left(\phi \right)= \int \d ^{d}x \left\lbrace\ud
\left[ \nabla \phi (x) \right]^{2}+\ud r\phi
^{2} (x)+\frac{1}{4!}g_0\Lambda^{4-d}\bigl(\phi^2(x))\bigr)^2 \right\rbrace , \eqnn $$
where $r$ is a regular function of the temperature $T$. The parameter $\Lambda$ has the dimension of a mass and corresponds to the inverse microscopic scale. It also appears as a cut-off in the Feynman diagrams of the perturbative expansion.
\par
For some value $r_c=r(T_c)$ the correlation length $\xi$ diverges (the physical mass
$m=1/\xi$ vanishes).  Near the critical temperature $T_c$, in  the critical domain $|T-T_c|\propto |r-r_c|\ll \Lambda^2$, $|\left< \phi\right>|\ll \Lambda^{(d-2)/2}$, the mass remains small, $m\ll \Lambda $. In this limit a universal behaviour of thermodynamic quantities is expected. The study of the critical domain thus reduces to the study of the large
cut-off behaviour, i.e.~to renormalization theory and the corresponding
renormalization group. However, in the traditional presentation of quantum field theory in  the context of particle physics, the dependence of the parameters of the action as a function of the cut-off $\Lambda$ is determined by the condition that renormalized correlation functions should have a finite cut-off limit. Here instead  the coefficient $g_0\Lambda^{4-d}/4!$ of the $\phi^4$ interaction has a dependence on $\Lambda$ given {\it a priori}. In particular in the dimensions of interest, $d<4$, the ``bare" coupling constant diverges for $\Lambda \to \infty$, a reflection of the IR instability of the gaussian fixed point, though the field theory, being
super-renormalizable, requires only a mass renormalization.
\nref\rWilFish{K.G. Wilson and M.E. Fisher, {\it Phys. Rev.  Lett.} 28 (1972)
240.}
\nref\rRGzemass{
C. Di Castro, {\it Lett. Nuovo Cimento}. 5 (1972) 69\semi
G. Mack, {\it Kaiserslautern 1972}, Lecture Notes in Physics  vol. 17, W.
Ruhl and A. Vancura eds. (Springer Verlag, Berlin 1972)\semi
B. Schroer, {\it Phys. Rev.} B8 (1973) 4200.}
\nref\rRGhomo{E. Br\'ezin, J.C. Le Guillou and J. Zinn-Justin, {\it Phys. Rev.} D8 (1973) 2418\semi
J. Zinn-Justin, {\it Carg\`ese Lectures 1973}, unpublished, later incorporated in ref.~\rBLGZJ\semi
C. Di Castro, G. Jona-Lasinio and L. Peliti, {\it Ann. Phys. (NY)} 87 (1974)
327.}
\nref\rCSBLGZJ{E. Br\'ezin, J.C. Le Guillou and J. Zinn-Justin, {\it Phys. Rev.} D8 (1973) 434\semi
P.K. Mitter, {\it Phys. Rev.} D7 (1973) 2927.}
\medskip
{\it The divergence of the bare coupling constant.} 
One solution to the problem of the large coupling constant  is provided  by Wilson--Fisher's famous $\varepsilon$-expansion \rWilFish. One defines, at least in perturbation theory, the field theory in arbitrary complex space dimension $d$. Setting $d=4-\varepsilon$ one then expands both in $g_0$ and $\varepsilon$. The large cut-off divergences become logarithmic as in four dimensions, and can be removed by standard renormalizations. 
Renormalization group (RG) equations for correlation functions follow, from which scaling laws can be derived, and which lead to calculations of physical quantities as power series in $\varepsilon$. \par
The most convenient RG equations are homogeneous differential equations, first derived for the critical (massless) theory \rRGzemass, and then generalized to the whole critical domain  $r\ne r_c$ \rRGhomo. They correspond to a scheme in which  the massless (or critical) theory has a perturbative expansion in $g_0$ and therefore rely on the double series expansion in $g_0$ and $\varepsilon$.  An alternative formalism is based on the Callan--Symanzik (CS) (inhomogeneous) equations for the massive theory \rCSBLGZJ. One introduces correlation functions for a renormalized field $\phi_\r=Z^{-1/2} \phi$, expressed in terms of the physical mass $m$ or inverse correlation length $\xi^{-1}=m$ and a renormalized coupling constant $g$.
They are implicitly defined by the conditions (after factorization of the obvious group indices):  
\eqna\egrzerom
$$ \eqalignno{\tilde \Gamma^{(2)}_{\rm R} (p;m ,g) & = m^2 +p^2 + O \left(p^4
\right), & \egrzerom{a} \cr\tilde \Gamma^{(4)}_{\rm R} \left(p_i=0;m ,g \right) & =
m^{4-d} g\,, & \egrzerom{b}  \cr} $$
where the functions $\tilde \Gamma^{(n)}_{\rm R}$ are the Fourier transforms of so-called (1PI) correlation functions $\Gamma^{(n)}_{\rm R}$, the coefficients of the expansion of the thermodynamic potential in terms of the local field expectation value $\langle  \phi_\r(x)\rangle $.\par
To prove scaling laws  from the CS equations it remains necessary to expand in $g$ and $\varepsilon$. The form of the CS $\beta$-function in  dimension $d$ is
$$\beta(g)=-(4-d)g+a_2(d)g^2+\cdots\  , \quad a_2(d)>0\,.$$
For $\varepsilon=4-d$ small it has a zero $g^*$ of order $\varepsilon$ and
one recovers the principle of the 
$\varepsilon$-expansion.\par
However  in this framework, because the theory now is massive, the perturbative expansion is IR finite in any dimension and the CS equations are always satisfied. Moreover one generally assumes that the results established for the $\varepsilon$-expansion remain valid at finite $\varepsilon$. Therefore  Parisi \ref\rParisi{G. Parisi, {\it Carg\`ese
Lectures 1973}, published in {\it J. Stat. Phys.} 23 (1980) 49.} has suggested 
working at fixed dimension $d<4$, in the massive theory (the massless theory is IR divergent). In contrast with the $\varepsilon$-expansion however, at fixed dimensions three or two {\it no small parameter}\/ is available. Therefore a precise determination of $g^*$ and all other physical quantities 
depends on the analytic properties of the series, in addition to the number of
terms available. A semi-classical analysis, based on instanton calculus,
unfortunately indicates that perturbation in quantum field theory is always
divergent.  Therefore to extract any information from perturbation theory a
summation method is required. Note finally that at fixed dimension, at any finite order,  universal quantities are given by expressions which depend on the renormalization scheme, in contrast
with the results of the $\varepsilon$-expansion.\par
This approach, which is more cumbersome, has eventually been tried for practical reasons: it is easier to calculate Feynman diagrams in dimension three than in
generic dimensions, and thus more perturbative orders could be obtained.\par 
While this approach is a natural extension of the $\varepsilon$-expansion, its interpretation directly at fixed dimension is worth discussing. 
At fixed dimension $d<4$, in the critical domain (condition which implies shifting $r$ by $r_c$ and thus performing a {\it mass renormalization}), all terms in the  perturbative expansion have a finite large cut-off limit at $u_0=g_0\Lambda^{4-d}$ fixed because the theory is super-renormalizable. This means physically that  the initial parameters of the field theory are
first tuned to remain artificially close to the unstable $u_0=0$ gaussian
fixed point. Indeed the true expansion parameter is a dimensionless quantity and therefore is proportional to $g_0\Lambda^{4-d}/m^{4-d}$.  To keep this parameter constant when $m$ the physical mass goes to zero for $T\to T_c$ ($r\to r_c$) in the high temperature phase, one must vary the dimensionless parameter $g_0$ as  
$$g_0^{1/(4-d)} \propto m/\Lambda\propto 1/\xi\Lambda\,.$$
The relevant theory, however, corresponds to the infinite $u_0$ limit.
One is then confronted with a serious technical problem: perturbation theory
is finite in the critical domain but one is interested in the infinite 
coupling limit, where obviously the perturbative expansion is no longer
useful. \par
One thus introduces the  renormalized coupling constant $g$ and the field renormalization defined by the conditions \egrzerom{}. Then
$$u_0 =g_0\Lambda^{(4-d)}=m^{(4-d)} G(g) , \quad \beta(g)=(d-4)G(g)/G'(g).\eqnd\euzegr $$
When the initial coupling constant $u_0$ becomes large the new dimensionless coupling $g$ has a finite limit provided the $\beta$-function  has an IR stable zero $g^*$: 
$$u_0\to\infty\ \Rightarrow\ \beta(g^*)=0\quad {\rm and}\
\omega\equiv\beta'(g^*)>0\,.$$ 
In such a situation the renormalized coupling $g$ is a more suitable expansion parameter then $u_0$.\par
The relations \euzegr~then imply that at $g_0$ fixed, $\Lambda/m\to
\infty$, 
$$g-g^*\propto (m/\Lambda)^\omega . \eqnd\eggstar $$
To the field renormalization $Z(g)$ is associated the RG function $\eta(g)$ $$\eta(g)=\beta(g)Z'(g)/Z(g)   .\eqnd\eZCSg $$
Integration of the equation \eZCSg~yields  the behaviour of $Z(g)$ near $g^*$
$$Z(g)\propto (g^*-g)^{\eta/\omega}\propto (m/\Lambda)^\eta, \eqnd\eZCSgstar
$$ where the exponent $\eta=\eta(g^*)$ characterizes the field anomalous
dimension $d_\phi$: $2d_\phi=d-2+\eta$. This singular behaviour, consistent with the scaling properties derived from the $\varepsilon$-expansion, explains why even in dimension $d<4$
the introduction of a field renormalization is necessary. 
\par
The approach has sometimes be questioned, because it involves this double limit, but the final results and their comparison with other data have shown the consistency of the method.\par
A last remark: in this framework the mass parameter $m$ has still to be related to the temperature. RG equations show that it is singular at $T_c$ and behaves for $t=r-r_c\to 0_+$ as 
$$m\propto t^\nu \propto (T-T_c)^\nu\,,\eqnd\enuexpo $$
 where $\nu$ is the
correlation length exponent. 
\nref\reviewLGZJ{A still partially
relevant reference is 
{\it Phase Transitions} vol. B72, M. L\'evy, J.C. Le Guillou and
J.~Zinn-Justin eds. Proceedings of the Carg\`ese Summer Institute 1980
(Plenum, New York 1982).}
\nref\rNMB{B.G.
Nickel, D.I.Meiron, G.B. Baker, Univ. of Guelph Report 1977\semi
see also 
G.A. Baker, B.G. Nickel, M.S. Green and D.I. Meiron, {\it Phys. Rev. Lett.} 36
(1976) 1351\semi G.A. Baker, B.G. Nickel, and D.I. Meiron, {\it Phys. Rev.} B17 (1978) 1365\semi
six-loop series have been published for generic values of $N$ by S.A. Antonenko and A.I. Sokolov, {\it Phys. Rev.} E51 (1995) 1894.}
\nref\rLGZJiii{J.C. Le Guillou
and J. Zinn-Justin, {\it J. Physique Lett. (Paris)} 46 (1985) L137; {\it J.
Physique (Paris)} 48 (1987) 19; {\it ibidem}\/ 50 (1989) 1365.}
\nref\repsexp{E. Br\'ezin, J.C. Le Guillou, J. Zinn-Justin and B.G. Nickel, {\it Phys. Lett.} 44A (1973) 227\semi
A.A. Vladimirov, D.I. Kazakov and O.V. Tarasov, {\it Zh. Eksp. Teor. Fiz.} 77
(1979) 1035 ({\it Sov. Phys. JETP} 50 (1979) 521)\semi
K.G. Chetyrkin, A.L. Kataev and F.V. Tkachov, {\it Phys. Lett.} B99 (1981)
147; B101 (1981) 457(E)\semi
K.G. Chetyrkin and F.V. Tkachov, {\it Nucl. Phys.} B192 (1981) 159\semi
K.G. Chetyrkin, S.G. Gorishny, S.A. Larin and F.V. Tkachov, {\it Phys. Lett.}
132B (1983) 351\semi
D.I. Kazakov, {\it Phys. Lett.} 133B (1983) 406\semi
S.G. Gorishny, S.A. Larin and F.V. Tkachov, {\it Phys. Lett.} 101A (1984)
120\semi
H. Kleinert, J. Neu, V. Schulte-Frohlinde, K.G. Chetyrkin and S.A.
Larin, {\it Phys. Lett.} B272 (1991) 39, {\it ibidem}\/ Erratum B319 (1993) 545.}

\section Critical exponents and series summation

Critical exponents are the most studied quantities in the theory of phase transitions, because they characterize universality classes and are easier to
calculate. They have been extensively used to test RG predictions by comparing
them with other results (experiments, high or low temperature
series expansion, Monte-Carlo simulations) \reviewLGZJ.  
The first precise determination of the exponents of the $O(N)$ symmetric
$N$-vector model has been reported by Le Guillou--Zinn-Justin \rLGZJ,  using the six-loop series for RG functions calculated by  Nickel {\it et al}\/ \rNMB. The summation methods used in \rLGZJ~were based on a Borel transformation and conformal
map. The same ideas have later been  applied to the
$\varepsilon$-expansion \rLGZJiii~when five loop series became available \repsexp, and
more recently to the equation of state by Guida--Zinn-Justin \rGuiZJ. With time the method has been refined
and the efficiency improved by various tricks but the basic principles have
not changed. 
\subsection Borel transformation and conformal mapping

Let $F(g)$ be any quantity given by a perturbative series
$$F(g)=\sum_{k=0} F_k g^k. \eqnn $$
A semi-classical analysis of barrier penetration effects for negative coupling (instantons) teaches us that in the $\phi^4_3$ field theory for large order $k$ the coefficients $F_k$  behave like
$$F_k \mathop{\propto}_{k\to \infty} k^s (-a)^k k!\,.$$
The value of $a>0$ has been  determined numerically (while $s$ is known
analytically). One then introduces $B(b,g)$, the Borel (rather Borel--Leroy)
transform of $F(g)$, which is defined by
$$B(b,g)=\sum_{k=0} {F_k \over \Gamma(b+k+1)} g^k, \eqnd\eBorseries $$
where $b$ is a free real parameter ($b>-1$). Formally, i.e.~in the sense of series
expansion, $F(g)$ can be recovered from
$$F(g)=\int_0^\infty t^b \e^{-t}B(b,gt)\d t\,.\eqnd\eBorLer $$
Using the large order estimate in \eBorseries~one verifies that  
$B(b,z)$ is analytic at least in a circle of radius $1/a$ and its singularity
closest to the origin is located at $z=-1/a$. Therefore $B(b,z)$, in
contrast with $F(z)$, is determined by its series expansion. However, 
the relation \eBorLer~can be extended from a relation between formal series to a relation between functions only if  $B(b,z)$ is analytic in a neighbourhood of the real positive axis. In the case of the $\phi^4_3$ field theory such a property has
been proven in constructive field theory (as well as the property that the
function $F(g)$ is indeed by given by \eBorLer) \ref\rBorsom{J.P. Eckmann, J.
Magnen and R. S\'en\'eor, {\it Commun. Math. Phys.} 39 (1975) 251\semi
J.S. Feldman and K. Osterwalder, {\it Ann. Phys. (NY)} 97 (1976) 80\semi
J. Magnen and R. S\'en\'eor, {\it Commun. Math. Phys.} 56 (1977) 237\semi
J.-P. Eckmann and H. Epstein,  {\it Commun. Math. Phys.} 68 (1979) 245.}.
Moreover it is necessary to construct an analytic continuation of the function $B(b,z)$ from the circle to the real positive axis. 
Considerations about more general instanton contributions strongly suggest that
the Borel transform actually is analytic in a cut-plane, the cut being on the
real negative axis, at the left of $-1/a$. Therefore an analytic continuation is provided by a conformal map of the cut-plane onto a circle:
$$z\mapsto u(z)={ \sqrt{1+az}- 1\over \sqrt{1+za}+ 1} \,.\eqnn $$
The function $B[b,z(u)]$ is then given by a series in powers of $u$ convergent in the cut-plane. 
The corresponding  (hopefully convergent) series expansion for function $F(g)$ takes the form  $$ F (g)= \sum_{k=0}^{\infty} U_{k}(b)
\int^{\infty}_{0}t^b\e^{-t} \left[ u (gt)
\right]^{k} \d t\,.\eqnn $$
The parameter $b$, as well as a few other parameters introduced in variants,
are used to improve the apparent convergence and test the sensitivity of
results to their variations. Moreover the value of $b$ has to stay within a
reasonable range around the value $s$ predicted by the large order
behaviour. Finally the summation method  is expected to be
efficient mainly when the available coefficients $F_k$  behave already as predicted by the asymptotic large order estimate. 
\nref\FeMoW{M. Ferer, M.A. Moore and M. Wortis, {\it Phys. Rev.} B8 (1973) 5205.}
\nref\rNickel{B.G. Nickel, {\it Physica} 106A (1981) 48.}
\nref\rZJhts{J. Zinn-Justin, {\it J. Physique (Paris)} 42 (1981) 783.}
\nref\rAdler{J. Adler, {\it J. Phys.} A16 (1983) 3585.}
\nref\rFiCh{J.H. Chen, M.E. Fisher and B.G. Nickel, {\it Phys. Rev. Lett.}
48 (1982) 630\semi
M.E. Fisher and J.H. Chen, {\it J. Physique (Paris)} 46 (1985) 1645.}
\nref\RiFiAl{D.S. Ritchie and M.E. Fisher, {\it Phys. Rev.} B5 (1972) 2668\semi
S. McKenzie, C. Domb and D.L. Hunter, {\it J. Phys.} A15 (1982) 3899\semi
 M. Ferer and A. Hamid-Aidinejad, {\it Phys. Rev.} B34 (1986) 6481.}  
\nref\rNiRe{B.G. Nickel and J.J. Rehr, {\it J. Stat. Phys.} 61 (1990) 1.}
\nref\rMacDo{D. MacDonald, D.L. Hunter, K. Kelly and N. Jan, {\it J. Phys.} A25
(1992) 1429.} 
\nref\rGutt{A.J.~Guttmann, {\it J. Phys.} A20 (1987) 1855\semi
A.J.~Guttmann and I.G.~Enting, {\it J. Phys.} A27 (1994) 8007.} 
\nref\rBCGS{G. Bhanot, M.
Creutz,  U. Gl\"assner and K. Schilling, {\it Phys. Rev.} B49 (1994) 12909.} 
\nref\rBUCOM{P. Butera and M. Comi, {\it Phys. Rev.} B56 (1997) 8212,
hep-lat/9703018.} 
\nref\rCPRV{M. Campostrini, A. Pelissetto, P. Rossi, E. Vicari, {\it Phys. Rev.} E60 (1999) 3526, cond-mat/9905078.} 
\nref\rJanke{W. Janke, {\it Phys.Lett.} A148 (1990) 306.}
\nref\rFeLa{A.M. Ferrenberg and D.P. Landau, {\it Phys. Rev.} B44 (1991)
5081.} 
\nref\rRaGu{C.F. Baillie, R.
Gupta, K.A. Hawick and G.S. Pawley, {\it Phys. Rev.} B45 (1992) 10438,
and references therein\semi 
R. Gupta and P. Tamayo, {\it Int. J. Mod. Phys.} C7 (1996) 305, cond-mat/9601048.} 
\nref\rJAHO{C. Holm and W. Janke, {\it Phys. Rev.} B48 (1993) 936;
{\it Phys. Lett.} A173 (1993) 8, hep-lat/9605024\semi
K. Chen, A.M. Ferrenberg and D.P. Landau, {\it Phys. Rev.} B48 (1993) 3249;
{\it J. Appl. Phys.} 73 (1993) 5488.
}
\nref\rsokal{B. Li, N. Madras and A.D. Sokal, {\it J. Stat. Phys.}  80 (1995) 661. }
\nref\rKAKA{K. Kanaya, S. Kaya, {\it Phys. Rev.} D51 (1995) 2404.}
\nref\rBloet{H.W.J. Bl\"ote, E. Luijten, J.R. Heringa, {\it   J. Phys.} A28 (1995) 6289, cond-mat/9509016\semi
A.L. Talapov and H.W.J. Bl\"ote, {\it J. Phys. } A29 (1996) 5727, cond-mat/9603013.}
\nref\rBFMM{H.G. Ballesteros, L.A. Fernandez, V. Martin-Mayor and A. Munoz
Sudupe, {\it Phys. Lett.} B387 (1996) 125, cond-mat/9606203.
Values for $N=1$ from private communication of L.A.Fernandez.} 
\nref\rBeloNick{P. Belohorec and B.G. Nickel, {\it Accurate universal and
two-parameter model results from a Monte-Carlo renormalization group study}, Guelph Univ. preprint 27/09/97.}
\nref\rcacape{S. Caracciolo, M.S. Causo and  A. Pelissetto, {\it Phys. Rev.} E57 (1998) R1215, cond-mat 9703250.}
\nref\rHaPiVi{M. Hasenbusch, K. Pinn, S. Vinti, {\it Phys. Rev.} B59 (1999) 11471, hep-lat 
/9806012\semi 
M. Hasenbusch and T. Török,  {\it J. Phys. A} 32 (1999) 6361, cond-mat/9904408.}

\subsection Exponents

The values of critical exponents obtained from field theory have remained
after about twenty years among the most precise determinations. Only recently
have consistent, but significantly more precise, experimental results  been
reported in low gravity superfluid experiments \ref\rLipa{ J.A. Lipa, D.R.
Swanson, J. Nissen, T.C.P. Chui and U.E. Israelson, {\it Phys. Rev. Lett.} 76
(1996) 944.}. Also the precision of results coming from high temperature expansions \refs{\FeMoW{--}\rCPRV} and various numerical simulations \refs{\rJanke{--}\rHaPiVi} on the lattice has kept steadily improving.\par 
Recently seven-loop terms have been obtained for $0\le N\le  3$
for two of the three RG functions, related to the $\phi$ and $\phi^2$ dimensions, by Murray and Nickel \ref\rBGNMU{D.B. Murray and B.G. Nickel, unpublished  Guelph report, 1991.}.
These terms, together with some improvement in the summation methods, have led to the new slightly more precise values of $g^*$ and critical exponents  displayed in table \label{\ttdiii} ($\g_{Ni}^*=(N+8)g^*/48\pi$) (Guida--Zinn-Justin \rGuiZJii).
Among the exponents $\gamma, \nu, \eta, \beta, \alpha $, only two are independent, for example
$$\gamma=\nu(2-\eta),\quad \beta =\ud \nu(1+\eta),\quad \alpha =2-3\nu\,,$$
but they are calculated independently to test the precision of the summation method.\par
The main
improvements concern the exponent $\eta$ which was poorly
determined, and the lower value of $\gamma$ for $N=0$ (polymers). 
\topinsert
\tablbl{\ttdiii}
$$ \vbox{\elevenpoint\offinterlineskip\tabskip=0pt\halign to \hsize
{& \vrule#\tabskip=0em plus1em & \strut\ # \
& \vrule#& \strut # 
& \vrule#& \strut # 
& \vrule#& \strut # 
& \vrule#& \strut # 
&\vrule#\tabskip=0pt\cr
\noalign{ \centerline{Table \ttdiii}\tableskip}
\noalign{\centerline{\it Critical exponents of the
$O(N)$ models from $d=3$ expansion  \rGuiZJii.} \tableskip}
\fileth
height2.0pt& \omit&& \omit&& \omit&&\omit&& \omit&\cr
&$ \hfill N \hfill$&&$ \hfill 0
\hfill$&&$ \hfill 1 \hfill$&&$ \hfill 2
\hfill$&&$ \hfill 3 \hfill$&\cr
height2.0pt& \omit&& \omit&& \omit&& \omit&& \omit&\cr
\fileth
height2.0pt& \omit&& \omit&& \omit&& \omit&& \omit&\cr
&$ \hfill \g^*_{\rm Ni} \hfill$
&&$ \hfill 1.413\pm 0.006  \hfill
$&&$ \hfill 1.411\pm0.004\hfill$
&&$ \hfill 1.403\pm 0.003 \hfill$
&&$\hfill 1.390\pm0.004 \hfill$&\cr
height2.0pt& \omit&& \omit&& \omit&& \omit&& \omit&\cr
&$ \hfill g^* \hfill$
&&$ \hfill  26.63\pm 0.11  \hfill$
&&$ \hfill 23.64\pm0.07 \hfill$
&&$ \hfill 21.16\pm 0.05 \hfill$
&&$\hfill 19.06\pm0.05 \hfill$&\cr 
height2.0pt& \omit&& \omit&& \omit&& \omit&& \omit&\cr
&$ \hfill \gamma \hfill$
&&$ \hfill 1.1596\pm0.0020 \hfill$
&&$ \hfill 1.2396\pm 0.0013\hfill$
&&$ \hfill 1.3169\pm0.0020  \hfill$
&&$\hfill 1.3895\pm0.0050 \hfill$&\cr
height2.0pt& \omit&& \omit&& \omit&& \omit&& \omit&\cr
&$ \hfill \nu \hfill$
&&$ \hfill 0.5882\pm 0.0011 \hfill$
&&$ \hfill 0.6304\pm 0.0013\hfill$
&&$ \hfill 0.6703\pm 0.0015  \hfill$
&&$\hfill 0.7073\pm 0.0035 \hfill$&\cr
height2.0pt& \omit&& \omit&& \omit&& \omit&&\omit&\cr
&$ \hfill \eta \hfill$
&&$ \hfill 0.0284\pm0.0025\hfill$
&&$ \hfill 0.0335\pm0.0025\hfill$
&&$ \hfill 0.0354\pm0.0025  \hfill$
&&$\hfill 0.0355\pm0.0025 \hfill$&\cr
height2.0pt& \omit&& \omit&& \omit&& \omit&& \omit&\cr
&$ \hfill \beta \hfill$
&&$ \hfill 0.3024\pm0.0008\hfill$
&&$ \hfill 0.3258\pm0.0014\hfill$
&&$ \hfill 0.3470\pm0.0016  \hfill$
&&$\hfill 0.3662\pm0.0025 \hfill$&\cr
height2.0pt& \omit&& \omit&& \omit&& \omit&& \omit&\cr
&$ \hfill \alpha \hfill$&&
$ \hfill 0.235\pm0.003 \hfill$
&&$ \hfill 0.109\pm0.004\hfill$
&&$ \hfill -0.011\pm0.004  \hfill$
&&$\hfill -0.122\pm0.010 \hfill$&\cr
height2.0pt& \omit&& \omit&& \omit&& \omit&& \omit&\cr
&$ \hfill \omega \hfill$&
&$ \hfill 0.812\pm0.016  \hfill$&
&$ \hfill 0.799\pm0.011\hfill$&
&$ \hfill 0.789\pm 0.011 \hfill$&
&$\hfill 0.782\pm0.0013 \hfill$&\cr
height2.0pt& \omit&& \omit&& \omit&& \omit&& \omit&\cr
&$ \hfill \theta=\omega\nu \hfill$&
&$ \hfill 0.478\pm0.010  \hfill$&
&$ \hfill 0.504\pm0.008\hfill$&
&$ \hfill 0.529\pm0.009  \hfill$&
&$\hfill 0.553\pm0.012 \hfill$&
\cr
height2.0pt& \omit&& \omit&& \omit&& \omit&& \omit&\cr
\fileth }}$$
\endinsert
In the framework of the $\varepsilon$-expansion, since the series used earlier were affected by a small error at order $\varepsilon^5$, the values have also been updated (table \label{\tteps}). Two kinds of results are reported, free means simple summation as above, bc means that the known values in two dimensions have been incorporated in the summation procedure. 
It is gratifying that the overall consistency between the two set of values coming from 3D series and $\varepsilon$-expansion has improved.
\midinsert
\tablbl\tteps
$$ \vbox{\elevenpoint\offinterlineskip\tabskip=0pt\halign to \hsize
{& \vrule#\tabskip=0em plus1em & \strut\ # \
& \vrule#& \strut #
& \vrule#& \strut # 
& \vrule#& \strut # 
& \vrule#& \strut # 
&\vrule#\tabskip=0pt\cr
\noalign{\centerline{Table \tteps}\tableskip}
\noalign{\centerline{\it
Critical exponents of the
$O(N)$ models from $\varepsilon$-expansion \rGuiZJii.}
\tableskip}
\fileth
height2.0pt& \omit&& \omit&& \omit&&\omit&& \omit&\cr
&$ \hfill N \hfill$&&$ \hfill 0
\hfill$&&$ \hfill 1 \hfill$&&$ \hfill 2
\hfill$&&$ \hfill 3 \hfill$&\cr
height2.0pt& \omit&& \omit&& \omit&& \omit&& \omit&\cr
\fileth
height2.0pt& \omit&& \omit&& \omit&& \omit&& \omit&\cr
&
$ \hfill\textstyle \gamma\ (\free)\atop \textstyle\gamma\ (\bc) \hfill$&&
$ \hfill\textstyle 1.1575\pm0.0060 \atop \textstyle 1.1571\pm0.0030\hfill$&&
$ \hfill\textstyle 1.2355\pm 0.0050\atop \textstyle 1.2380\pm0.0050\hfill$&&
$ \hfill\textstyle 1.3110\pm0.0070\atop \textstyle 1.317 \hfill$&&
$\hfill \textstyle 1.3820\pm0.0090\atop\textstyle 1.392 \hfill$&\cr
height2.0pt& \omit&& \omit&& \omit&& \omit&& \omit&\cr
&$ \hfill \textstyle \nu\ (\free)\atop \textstyle\nu\ (\bc) \hfill$&&
$ \hfill\textstyle 0.5875\pm0.0025 \atop \textstyle 0.5878\pm0.0011\hfill$&&
$ \hfill\textstyle 0.6290\pm0.0025\atop\textstyle 0.6305\pm0.0025\hfill$&&
$ \hfill \textstyle 0.6680\pm0.0035 \atop \textstyle 0.671\hfill$&&
$\hfill\textstyle 0.7045\pm0.0055 \atop\textstyle 0.708\hfill$&\cr
height2.0pt& \omit&& \omit&& \omit&& \omit&&\omit&\cr
&$ \hfill \textstyle \eta\ (\free)\atop \textstyle\eta\ (\bc) \hfill$&&
$ \hfill\textstyle 0.0300\pm0.0050\atop\textstyle 0.0315\pm0.0035 \hfill$&&
$ \hfill\textstyle 0.0360\pm0.0050\atop\textstyle 0.0365\pm0.0050\hfill$&&
$ \hfill\textstyle 0.0380\pm0.0050 \atop\textstyle 0.0370 \hfill$&&
$\hfill\textstyle 0.0375\pm0.0045\atop\textstyle 0.0355 \hfill$&\cr
height2.0pt& \omit&& \omit&& \omit&& \omit&& \omit&\cr
&$ \hfill \textstyle \beta\ (\free)\atop \textstyle\beta\ (\bc) \hfill$&&
$ \hfill\textstyle 0.3025\pm0.0025\atop\textstyle 0.3032\pm0.0014 \hfill$&&
$ \hfill\textstyle 0.3257\pm0.0025\atop\textstyle 0.3265\pm0.0015\hfill$&&
$ \hfill 0.3465\pm0.0035  \hfill$&&
$\hfill 0.3655\pm0.0035 \hfill$&\cr
height2.0pt& \omit&& \omit&& \omit&& \omit&& \omit&\cr
&$ \hfill \omega \hfill$&&
$ \hfill 0.828\pm0.023  \hfill$&&
$ \hfill 0.814\pm0.018 \hfill$&&
$ \hfill 0.802\pm 0.018 \hfill$&&
$\hfill 0.794 \pm0.018 \hfill$&\cr
height2.0pt& \omit&& \omit&& \omit&& \omit&& \omit&\cr
&$ \hfill \theta \hfill$&&
$ \hfill 0.486\pm0.016  \hfill$&&
$ \hfill 0.512\pm0.013 \hfill$&&
$ \hfill 0.536\pm 0.015 \hfill$&&
$\hfill 0.559 \pm0.017 \hfill$&\cr
height2.0pt& \omit&& \omit&& \omit&& \omit&& \omit&\cr
\fileth}}$$
\endinsert
Finally $O(4)$ results of interest for simulations of the Higgs phase transition at finite temperature, obtained from six loop series, have been added (table \label{\ttniv}).
 \midinsert
\tablbl\ttniv
$$ \vbox{\elevenpoint\offinterlineskip\tabskip=0pt\halign to \hsize
{& \vrule#\tabskip=0em plus1em & \strut\ # \
& \vrule#& \strut #
& \vrule#& \strut # 
&\vrule#\tabskip=0pt\cr
\noalign{ \centerline{Table \ttniv}\tableskip}
\noalign{\centerline{\it  Critical exponents in the
$O(4)$ models from $d=3$ and $\varepsilon$-expansion \rGuiZJii.} \tableskip}
\fileth
height2.0pt& \omit&& \omit&& \omit&\cr
&$ \hfill \hfill$&&$ \hfill d=3\hfill$&&$ \hfill \varepsilon:\free,\bc
\hfill$&\cr 
height2.0pt& \omit&& \omit&& \omit&\cr
\fileth
height2.0pt& \omit&& \omit&& \omit&\cr
&$ \hfill \g^*_{\rm Ni} \hfill$
&&$ \hfill 1.377\pm 0.005  \hfill$
&&$ \hfill\hfill$&\cr
height2.0pt& \omit&& \omit&& \omit&\cr
&$ \hfill g^* \hfill$
&&$ \hfill 17.30\pm 0.06  \hfill$
&&$ \hfill \hfill$
&\cr
height2.0pt& \omit&& \omit&& \omit&\cr
&$ \hfill \gamma \hfill$
&&$ \hfill 1.456\pm 0.010 \hfill$
&&$ \hfill 1.448\pm 0.015\ ,\ 1.460\hfill$&
\cr
height2.0pt& \omit&& \omit&& \omit&\cr
&$ \hfill \nu \hfill$
&&$ \hfill 0.741\pm 0.006  \hfill$
&&$ \hfill 0.737\pm 0.008\ , \ 0.742 \hfill$&
\cr
height2.0pt& \omit&& \omit&& \omit&\cr
&$ \hfill \eta \hfill$
&&$ \hfill 0.0350\pm0.0045\hfill$
&&$ \hfill 0.036\pm 0.004\ , \ \ 0.033 \hfill$&
\cr
height2.0pt& \omit&& \omit&& \omit&\cr
&$ \hfill \beta \hfill$
&&$ \hfill 0.3830\pm 0.0045\hfill$
&&$ \hfill 0.3820\pm 0.0025\hfill$
&\cr
height2.0pt& \omit&& \omit&& \omit&\cr
&$ \hfill \alpha \hfill$
&&$ \hfill -0.223\pm 0.018 \hfill$
&&$ \hfill -0.211\pm 0.024 \hfill$&\cr
height2.0pt& \omit&& \omit&& \omit&\cr
&$ \hfill \omega \hfill$
&&$ \hfill 0.774\pm 0.020  \hfill$
&&$ \hfill 0.795\pm 0.030\hfill$
&\cr
height2.0pt& \omit&& \omit&& \omit&\cr
&$ \hfill \theta \hfill$&
&$ \hfill0.574\pm 0.020\hfill$&
&$ \hfill 0.586\pm 0.028\hfill$&\cr
height2.0pt& \omit&& \omit&& \omit&\cr
\fileth }}$$
\endinsert
\section The scaling equation of state 

Let us first recall a few properties of the equation of the state in
the critical domain, in the specific case $N=1$ (Ising-like
systems), at $d<4$.\sslbl\ssscaleqst 
\par 
The equation of state is the relation between 
magnetic field $H$, magnetization $M=\left<\phi\right>$ (the ``bare"
field expectation value) and the temperature which is represented by the
parameter $t= r-r_c\propto T-T_c$. It is related to
the free energy per unit volume, in field theory
language the generating functional $\Gamma(\phi)$ of 1PI correlation
functions restricted to constant fields, i.e~the effective potential $\cal V$,
${\cal V}(M)=\Gamma(M)/{\rm vol.}$, by $H=\del {\cal V} /\del M$.
In the critical domain the equation of state has Widom's scaling form
($\delta=(d+2-\eta)/(d-2+\eta)$)
$$H(M,t)=M^\delta f(t/M^{1/\beta}), \eqnd\ehscal $$ 
a form initially conjectured and which renormalization group has justified. 
\par
One property of the function $H(M,t)$ which plays an essential 
role in the analysis is {\it Griffith's analyticity}: 
it is regular at $t=0$ for $M>0$ fixed, and simultaneously it is regular at
$M=0$  for $t>0$ fixed. 
\medskip
{\it Amplitude ratios.} 
Universal amplitude ratios are numbers characterizing the behaviour of 
thermodynamical quantities near $T_c$. Several of them commonly considered in
the literature can be directly derived from the scaling equation of state. Let
us just give two examples.\par
The singular part of the specific heat,
i.e.~the $\phi^2$ two-point correlation function at zero momentum, behaves
like 
$$C_H=A^{\pm}\left\vert t \right\vert^{-\alpha},\qquad t\propto T-T_c
\rightarrow \pm 0\,. \eqnn $$
The ratio $A^+/A^-$ then is universal.\par
The magnetic susceptibility $\chi$ in
zero field, i.e.~the $\phi$ two-point function at zero momentum, diverges like 
$$\chi = C^{\pm}\left\vert t \right\vert^{-\gamma}, \qquad t \rightarrow \pm 0
\, . \eqnn $$
The ratio $C^+/C^-$ is also universal.
\medskip
{\it The $\varepsilon$-expansion.} 
The first results concerning the scaling equation of state
have been obtained within the framework of the $\varepsilon=4-d$
expansion. \par
The $\varepsilon$-expansion of the scaling equation of state 
has been determined up to order $\varepsilon^2$ for the general $O(N)$ model
\ref\rBWW{G.M. Avdeeva and A.A. Migdal, {\it JETP Lett.} 16 (1972) 178\semi
E. Br\'ezin, D.J. Wallace and K.G. Wilson, {\it Phys. Rev. Lett.} 29 (1972)
591; {\it Phys. Rev.} B7 (1973) 232.}, and order $\varepsilon^3$ for $N=1$
\ref\rWZAN{D.J. Wallace and R.P.K. Zia, {\it J.
Phys.} C7 (1974) 3480.}. We give here the function $f(x=t/M^{1/\beta})$ of eq.~\ehscal~for $N=1$ up to order $\varepsilon^2$ to display its structure,
$$f(x)=1+x+\frac{1}{6}\varepsilon (x+3) L+\varepsilon^2 \left[
\frac{1}{72}(x+9)L^2+\frac{25}{324}(x+3)L\right]+
O\left(\varepsilon^3\right)  ,\eqnd\estateps $$
with $L=\log(x+3)$.\par
The expression \estateps~is not valid for $x$ large, i.e.~for small
magnetization $M$. In this regime the magnetic field $H$ has a regular
expansion in odd powers of $M$. It is thus convenient to express the equation
of state in terms of another scaling variable $z \propto x^{-\beta}$ because
$$z\propto M t^{-\beta}.\eqnd\ezscalvar $$ 
The equation of state then takes the form
$$H\propto t^{\beta\delta}F(z) ,\eqnd\estatmag $$ 
where the relation between exponents $\gamma=\beta(\delta-1)$ has been used.
Substituting into eq.~\estateps~$x=x_0 z^{-1/\beta}$ (the constant $x_0$
takes care of the normalization of $z$) and expanding in
$\varepsilon$  one finds at order $\varepsilon^2$ for the function \estatmag
$$F(z)=\F_0(z)+\varepsilon \F_1(z)+\varepsilon^2 \F_2(z)+
O(\varepsilon^3),$$
with
\def\L{\tilde L}
\eqna\estateii
$$ \eqalignno{\F_0& = z + \frac{1}{6}z^3 &\estateii{a}\cr
\F_1& =\frac{1}{12}\bigl(-z^3 + \L( 2z + z^3 )\bigr)\cr
\F_2& = \frac{1}{1296}\bigl(-50z^3 + \L( 100z - 4z^3 ) +\L^2( 18z +
27z^3)\bigr) &\estateii{b}\cr }$$
and $\L=\log(1+z^2/2)$.\par
Within the framework of the formal $\varepsilon$-expansion one can
easily pass from one expansion to the other. Still a matching problem
arises if one wants to use the $\varepsilon$-expansion to determine the
equation of state for $d=3$, i.e.~$\varepsilon=1$. One is thus naturally led
to look for a uniform representation valid in both limits. 
Josephson--Schofield parametric representation \ref\rJoseph{P.
Schofield, {\it Phys. Rev. Lett.} 23 (1969) 109\semi P. Schofield, J.D.
Litster and J.T. Ho, {\it Phys. Rev. Lett.} 23 (1969) 1098\semi B.D.
Josephson, {\it J. Phys.} C2 (1969) 1113.}~has this
property.   
\section Parametric representation of the equation of state

In order to implement both Griffith's analyticity and the scaling relation, one parametrizes the equation of state in terms of two new variables $R$ and
$\theta$, setting: 
$$\left\lbrace\eqalign{M &= m_0 R^{\beta}\theta\, , \cr t& = R
\left(1-\theta^2 \right),  \cr H & =h_0 R^{\beta\delta}h(\theta)\, ,
\cr}\right. \eqnn $$
where $h_0, m_0$ are two normalization constants. We choose $h_0$ such that 
$$ h(\theta)=\theta+O\left(\theta^3\right)\, .$$
In terms of the scaling variables $x$ of eq.~\estateps{} or $z$
from eq.~\ezscalvar~this parametrization corresponds to set 
$$\eqalignno{z&=\rho \theta/ (1-\theta^2)^\beta,\quad \theta >
0\,, & \eqnd\ezmaptheta \cr
x&=x_0\rho^{-1/\beta}\left(1-\theta^2\right)\theta^{-1/\beta} ,&\eqnn \cr
}$$
where $\rho$ is some other positive constant. \par 
Then the function $h(\theta)$   
is an odd function of $\theta$ which from Griffith's analyticity is regular
near $\theta =1$, which is $x$ small, and near $\theta=0$ which is $x$ large. 
It vanishes for $\theta=\theta_0$ which corresponds to the coexistence curve
$H=0,T<T_c$. \par 
Note that the mapping \ezmaptheta~is not invertible for values of $\theta$
such that $z'(\theta)=0$. The derivative vanishes for
$\theta=1/\sqrt{(1-2\beta)}\approx 1.69$. One has to verify that this value is 
reasonably larger than $\theta_0$, the largest possible value of $\theta$.  
\par
Finally it is useful for later purpose to write more explicitly the relation
between the function $F(z)$ of eq.~\estatmag~and the function $h(\theta)$:
$$h(\theta)=\rho^{-1}
\left(1-\theta^2\right)^{\beta\delta}F\left(z(\theta)\right). \eqnd\emagfparii
$$ 
Expanding both functions
$$\eqalignno{F(z)&=z+\frac{1}{6}z^3+\sum_{l=2}F_{2l+1}z^{2l+1},&\eqnn \cr
h(\theta)/\theta&=1+\sum_{l=1}h_{2l+1} \theta^{2l} ,&\eqnn \cr}$$
one finds the relations
\eqna\eFHrelat
$$\eqalignno{h_3&=\frac{1}{6}\rho^2-\gamma & \eFHrelat{a}\cr
h_5&=\ud\gamma(\gamma-1)+\frac{1}{6}(2\beta-\gamma)\rho^2+F_5\rho^4 & 
\eFHrelat{b}\cr
h_7&=\frac{1}{6}\gamma(\gamma-1)(\gamma-2)+\frac{1}{12}(2\beta-\gamma)
(2\beta-\gamma+1)\rho^2\cr&\quad +(4\beta-\gamma)F_5\rho^4 +F_7\rho^6
&\eFHrelat{c} 
\cr &\cdots \cr}$$
From the parametric representation of the equation of state it is then
possible to derive a representation for the singular part of the free energy
per unit volume as well as various universal ratios of amplitudes.
\nref\rNA{J.F. Nicoll and P.C. Albright, {\it Phys. Rev.} B31
(1985) 4576.} 
\medskip
{\it Parametric representation and $\varepsilon$-expansion.} 
Up to order $\varepsilon^2$ the constant $m_0$ (or $\rho$) can be chosen in
such a way that the function $h(\theta)$ reduces to:  
$$ h(\theta)= \theta\left(1 -\frac{2}{3}\theta^2 \right) +O\left(\varepsilon^2
\right). \eqnn $$
The minimal model in which $h(\theta)$ is approximated by a cubic odd function
of $\theta$ is called the linear parametric model. At order $\varepsilon^2$
the linear parametric model is exact, but at order $\varepsilon^3$ the
introduction of a term proportional to $\theta^5$ becomes necessary
\refs{\rWZAN,\rNA}. One finds:   
$$h(\theta)=\theta (1+h_3\theta^2+h_5\theta^4)+O\left(\varepsilon^4\right),
\eqnn $$ 
with
$$h_3=-{2\over3}\left(1+{\varepsilon^2 \over 12} \right) ,\quad
h_5={\varepsilon^3 \over 27}\left(\zeta(3)-\ud\lambda -
\frac{1}{4}\right),\eqnn $$  
where $\lambda$ is the constant 
$$\lambda  =\frac{1}{3}\psi'(1/3)-\frac{2}{9}\pi^2=1.17195361934 \ldots\
. \eqnn $$ 
The function $h(\theta)$ vanishes on the coexistence curve for
$\theta=\theta_0$: 
$$\theta_0^2  ={3 \over 2}\left(1-{\varepsilon^2 \over 12} \right)
+O(\varepsilon^3) .
\eqnn $$ 
Note that $h_3$ and thus $\theta_0$  are determined only up to order
$\varepsilon^2$. 
It follows
$$\rho^2=6(\gamma+h_3)=2\left(1+\ud\varepsilon+\frac{7}{108}\varepsilon^2
\right)=3.13\pm0.13,$$ 
because $h_3$ is determined only up to order $\varepsilon^2$,
\medskip
{\it Remark.} In the more general $O(N)$ case, the  
parametric representation also automatically generates
equations of state with satisfy the required regularity properties, and thus
leads to uniform approximations. However for $N>1$ the function $h(\theta)$
still has a singularity on the coexistence curve, due to the presence of
Goldstone modes in the ordered phase and has therefore a more complicated
form. The nature of this singularity can be obtained from the study of the
non-linear $\sigma$-model. It is not clear whether a simple polynomial
approximation would be useful. For $N=1$ instead, one expects at most
an essential singularity on the coexistence curve, due to barrier penetration,
which is much weaker and non-perturbative in the small $\varepsilon$- or small
$g$-expansion.
\section Perturbative expansion at fixed dimension three

We now discuss the calculations based on a perturbative
expansion at fixed dimension $d=3$. Five loop series for the renormalized
effective potential of the $\phi^4_3$ theory have been first reported in
Bagnuls {\it et al.}~\ref\rBBMN{C. Bagnuls, C. Bervillier, D.I. Meiron and
B.G. Nickel, {\it Phys. Rev.} B35 (1987) 3585.},
but the printed tables contain some serious misprints. These  have been
noticed by Halfkann and Dohm who have published corrected values
\ref\rHaDo{F.J. Halfkann and V. Dohm, {\it Z. Phys.} B89 (1992) 79.}.
These five-loop calculations have only been performed for $N=1$, because they are
much more difficult for $N\ne 1$ due to  the presence of two lengths, the
correlation lengths along the applied field and transverse to it.
\subsection General remarks

The general framework again is the massive theory renormalized
at zero momentum. The correlation functions $\Gamma^{(n)}_\r$ of the
renormalized field $\phi_\r =\phi/\sqrt{Z}$ are
fixed by the normalization conditions \egrzerom{}.
Eventually the renormalized  coupling constant $g$ has to be set to its IR
fixed point value $g^*$. \par  
The conditions \egrzerom{} imply that the free energy $\cal F$
expressed in terms of the ``renormalized" magnetization $\varphi$,  
i.e.~the expectation value of the renormalized field $\varphi=\left<\phi_\r
\right>$, has a small $\varphi$ expansion of the form (in $d$ dimensions) 
$${\cal F}(\varphi)={\cal F}(0)+\ud m^2
\varphi^2+\frac{1}{4!}m^{4-d} g \varphi^4 +O\left(\varphi^6\right). \eqnn $$
It is important to remember that the finite field renormalization $ Z(g)$ is singular at
$g^*$ (equation \eZCSgstar). It follows that $\varphi\propto M/m^{\eta/2}$.
\par
It is convenient to introduce the rescaled variable $z$
$$z=\varphi m^{(2-d)/2} \sqrt{g}\propto M/m^{(d-2+\eta)/2} \propto M/t^\beta\,,\eqnn $$
(equation \enuexpo) and set
$${\cal F}(\varphi)-{\cal F}(0)={m^d\over g}{\cal V}(z,g). \eqnn $$
Taking into account the definitions of section \ssscaleqst, we conclude
that the equation of state is related to the derivative $F$ of the reduced 
effective potential ${\cal V}$ with respect to $z$ 
$$F(z,g)={\partial {\cal V}(z,g)\over \partial z}\,,\eqnd\eVpotFeq $$
by
$$H\propto t^{\beta\delta}F(z) . $$
Ising symmetry implies that $F$ is an odd function of $z$
$$F(z,g)=z+\frac{1}{6}z^3+\sum_{l=2}F_{2l+1}(g) z^{2l+1}\,.\eqnn $$
\subsection The problem of the low temperature phase

To determine the equation of state in the whole physical range, or universal
ratios of amplitudes, a new problem arises. In this framework it is
more difficult to calculate physical quantities in the ordered phase because
the theory is parametrized in terms of the disordered phase correlation length
$\xi=m^{-1}\propto (T-T_c)^{-\nu}$ which is singular at $T_c$
(as well as all correlation functions normalized as in~\egrzerom{}). 
For example at one-loop order for $d=3$ the scaling 
function $F(z,g)$ (eq.~ \eVpotFeq) is given by:\sslbl\ssUQfixD 
$$\eqalignno{F(z,g)& = z +\frac{1}{6}z^3 -\frac{1}{8\pi}gz\left[\left(1+z^2/2
\right)^{1/2}-1-z^2/4\right]& \eqnd\eoneloop\cr 
&= z +\frac{1}{6}z^3+\frac{1}{256\pi}gz^5-{1\over2^{13}\pi}gz^7+O(z^9),\cr} $$
where the subtractions, due to the mass and coupling normalizations,
are determined by the conditions \egrzerom{}.
This expression is adequate for the description of the disordered phase. However when  $t$ goes to zero for fixed magnetization, i.e.~$m\rightarrow 0$ at $\varphi$ fixed,
then $z\to \infty$ as seen in eq.~\ezscalvar. Thus all terms in the loopwise expansion become 
singular. In this limit one knows from eq.~\ehscal~that the equation of state behaves like
$$H(M,t=0)\propto M^{\delta}\ \Rightarrow\ F(z)\propto
z^\delta.\eqnd\ettozero $$
\par
In the framework of the $\varepsilon$-expansion the scaling relations (and thus the
limiting behaviour~\ettozero) are exactly satisfied order by order.
Moreover the change to the  variable $x\propto z^{-1/\beta}$ (more
appropriate for the regime $t\rightarrow 0$) gives an expression for 
$f(x)\propto F(x^{-\beta}) x^{\beta\delta}$ that is explicitly regular in
$x=0$ 
(Griffith's analyticity): the singular powers of $\log
x$ induced by the change of variables cancel non trivially at each order,
leaving only regular corrections. 
\par
The situation changes when one deals with the perturbation theory in $d=3$
dimensions: scaling is not satisfied for generic values of $g$ but only at
$g^*$. Consequently scaling properties are not satisfied order by order in
$g$. In particular the change to the Widom function $f(x)$ will introduce 
singular powers of $x$ that violate Griffith's analyticity.
An analogous problem arises if one first sums the series at $g=g^*$
before changing to the variable $x$.  
In this case the singular contributions (in the form of powers of $x$) do not
cancel, as a result of unavoidable numerical summation errors.
\par
Several approaches have been  proposed to solve the problem of continuation to the
ordered phase. A rather powerful method, motivated by the results obtained
within the $\varepsilon$-expansion scheme, is based on the parametric
representation \rGuiZJ. 
\medskip
{\it Parametric representation and order dependent mapping (ODM).}
The problem that one faces is the following: to reach the ordered region $t<0$
one must cross the point $z=\infty$. However we know from Griffith's
analyticity that $F(z)z^{-\delta}$ is regular in the variable $z^{-1/\beta}$.
This property is naturally satisfied within the parametric representation.
One thus introduces an new variable $\theta$ and an auxiliary function
$h(\theta)$ defined as in \eqns{\ezmaptheta,\emagfparii}:  
the exact function $h(\theta)$ is regular near $\theta=1$ (i.e.~$z=\infty$) and up to the coexistence curve. \par
The approximate $h(\theta)$ that one obtains by summing perturbation theory at
fixed dimension, instead is still not regular, because the singular terms generated
by the change of parametrization~\ezmaptheta~at $\theta=1$ do not cancel exactly due
to summation errors. The last step involves a Taylor expansion of
the approximate expression of $h(\theta)$ around $\theta=0$ and a 
truncation of the expansion, to enforce regularity. A question
then arises, to which order in $\theta$ should one expand?
Since the coefficients of the $\theta$ expansion are in one to one
correspondence with the coefficients of the small $z$ expansion of the
function $F(z,g^*)$, the maximal power of $\theta$ in $h(\theta )$, 
should be equal to the maximal power of $z$ whose coefficient can be
determined with reasonable accuracy. Indeed although the small $z$ expansion
of $F(z)$ at each finite loop order in $g$ contains an infinite number of
terms, the evaluation of the coefficients of the higher powers of $z$ is
increasingly difficult. The reasons are twofold:\par
(i) The number of terms of the series in $g$ required to get a
precise estimate of $F_l$ increases with $l$  because the large order behaviour sets in later.\par
(ii) At any finite order in $g$ the function $F(z)$ has spurious singularities
in the complex $z$ plane  (see e.g.~eq.~\eoneloop, $z^2=-2$) that dominate  
the behaviour of the coefficients $F_l$ for $l$ large.  \par
In view of these difficulties one has to ensure the fastest possible
convergence of the small $\theta$ expansion. For this purpose one uses the arbitrary parameter $\rho$ in eq.~\ezmaptheta:
one determines $\rho$ by minimizing the last term in the truncated small
$\theta$ expansion, thus increasing the importance of small powers of $\theta$
which are more precisely calculated. This is nothing but the application to
this particular example of the series summation method based on ODM
\ref\rOMD{R. 
Seznec and J. Zinn-Justin, {\it J. Math. Phys.} 20 (1979) 1398\semi 
J.C. Le Guillou and J. Zinn-Justin, {\it Ann. Phys. (NY)} 147 (1983) 57\semi
R. Guida, K. Konishi and H. Suzuki, {\it Ann. Phys. (NY)} 241 (1995) 152; {\it ibidem}\/ 249
(1996) 109.}.
\nref\rZLFish{
S. Zinn, S.-N. Lai, and M.E. Fisher, {\it Phys. Rev. E} 54 (1996) 1176.}
\goodbreak
\section Numerical results

The calculation proceeds in two steps; first one determines the first coefficients of the small
field expansion, then one introduces the parametric representation.
\midinsert
\epsfysize=60mm
\epsfxsize=110mm
\centerline{\epsfbox{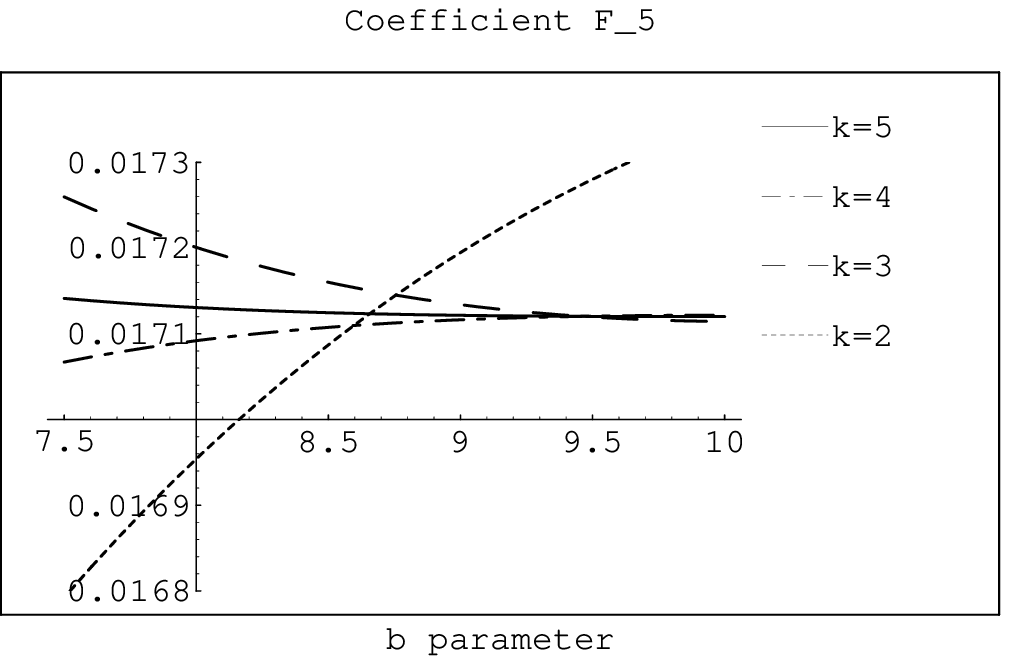}}
\figure{5.mm}{The summed coefficient $F_5$ as a function of the Borel--Leroy
parameter $b$ for successive orders $k$.} 
\figlbl\figmexi
\endinsert
\nref\rPelVic{A. Pelissetto and E. Vicari, {\it Nucl. Phys.} B519 (1998) 626, cond-mat/9711078;  
{\it ibidem}\/ B522 (1998) 605, cond-mat/9801098.}
\nref\rreisz{T. Reisz, {\it Phys. Lett.} B360 (1995) 77.}
\nref\rbuco{P. Butera and M. Comi, {\it Phys. Rev.} E55 (1997) 6391.}
\nref\rtsypin{M.M. Tsypin, {\it Phys. Rev. Lett.} 73 (1994) 2015.}
\nref\rwetterich{N. Tetradis and C. Wetterich, {\it Nucl. Phys.} 
B422 (1994) 541.}
\nref\rmorris {T. R. Morris, {\it Nucl. Phys.} B495 (1997) 477.}
\subsection The small field expansion

The determination of the coefficients $F_{2l+1}$ of the small
$z$ (small field) expansion of the function $F(z)$ relies on exactly the same method as for exponents, i.e.~Borel--Leroy
transformation and conformal map. In figure \figmexi~the behaviour of $F_5$ in terms of the Borel--Leroy parameter $b$
is displayed. Increasing flatness of the curves when $k$ increases,
i.e.~increasing insensitivity to the parameter $b$, supports the hypothesis that the 
method indeed converges.\par
Because the asymptotic regime sets in later when $l$ increases, the efficiency of the summation correspondingly decreases, as expected. Table \label{\tteqst} contains the results of \rGuiZJ~together with other published
estimates of the coefficients of the small $z$ expansion of $F(z)$ coming from high temperature series \refs{\rZLFish,\rreisz,\rbuco}, simulations \refs{\rtsypin} and   derivative expansion of the Exact RG \refs{\rwetterich,\rmorris}.
\midinsert
\tablbl\tteqst
$$ \vbox{\elevenpoint\offinterlineskip\tabskip=0pt\halign to \hsize
{& \vrule#\tabskip=0em plus1em & \strut\ # \ 
& \vrule#& \strut #
& \vrule#& \strut #  
& \vrule#& \strut #  
& \vrule#& \strut #  
&\vrule#\tabskip=0pt\cr
\noalign{\centerline{Table \tteqst} \tableskip}
\noalign{\centerline{\it Equation of state.}
\tableskip}
\fileth
height2.0pt& \omit&& \omit&& \omit&&\omit&& \omit&\cr
&$ \hfill  \hfill$&&$ \hfill g^* \hfill$&&$ \hfill
F_5  \hfill$&&$ \hfill F_7 \times 10^4
\hfill$&&$ \hfill F_9 \times 10^5\hfill$&\cr 
height2.0pt& \omit&& \omit&& \omit&& \omit&& \omit&\cr
\fileth
height2.0pt& \omit&& \omit&& \omit&& \omit&& \omit&\cr
&$ \hfill \varepsilon{-\rm exp.}, \refs{\rGuiZJ,\rGuiZJii} \hfill$
&&$ \hfill 23.3\hfill$
&&$ \hfill 0.0177\pm 0.0010 \hfill$
&&$ \hfill 4.8\pm0.6  \hfill$
&&$\hfill -3.3\pm0.3 \hfill$&\cr
height2.0pt& \omit&& \omit&& \omit&& \omit&& \omit&
\cr&$ \hfill \varepsilon{-\rm exp.}, \refs{\rPelVic} \hfill$
&&$ \hfill 23.4 \pm 0.1\hfill$
&&$ \hfill 0.01715\pm 0.00009 \hfill$
&&$ \hfill 4.9\pm0.6  \hfill$
&&$\hfill -5.5\pm 4 \hfill$&\cr
height2.0pt& \omit&& \omit&& \omit&& \omit&& \omit&\cr
&$ \hfill d=3,  \refs{\rGuiZJ,\rGuiZJii} \hfill$
&&$ \hfill 23.64\pm 0.07  \hfill$
&&$ \hfill0.01711\pm 0.00007 \hfill$
&&$ \hfill 4.9\pm 0.5   \hfill$
&&$\hfill -7\pm5 \hfill$&\cr 
height2.0pt& \omit&& \omit&& \omit&& \omit&& \omit&\cr
&\hfill  HT~\rZLFish \hfill
&&$ \hfill  24.45\pm0.15\hfill$
&&$ \hfill.017974\pm.00015\hfill$
&&$ \hfill  \hfill$
&&$ \hfill  \hfill$&\cr
height2.0pt& \omit&& \omit&& \omit&& \omit&&\omit&\cr
&\hfill   HT~\rreisz \hfill
&&$ \hfill 23.72 \pm 1.49 \hfill$
&&$ \hfill0.0205\pm 0.0052 \hfill$
&&$ \hfill   \hfill$
&&$ \hfill \hfill$&\cr
height2.0pt& \omit&& \omit&& \omit&& \omit&& \omit&\cr
&\hfill  HT~\rbuco \hfill
&&$ \hfill 23.69 \pm .10 \hfill$
&&$ \hfill.0168\pm 0.0012\hfill$
&&$ \hfill 5.4\pm 0.7 \hfill$
&&$ \hfill -2.3\pm 1.1  \hfill$&\cr
height2.0pt& \omit&& \omit&& \omit&& \omit&& \omit&\cr
&\hfill   MC~\rtsypin \hfill
&&$ \hfill 23.3 \pm 0.5  \hfill$
&&$ \hfill0.0227\pm0.0026 \hfill$
&&$ \hfill  \hfill$
&&$ \hfill  \hfill$&\cr
height2.0pt& \omit&& \omit&& \omit&& \omit&& \omit&\cr
&\hfill ERG~\rmorris \hfill
&&$ \hfill 20.72\pm 0.01  \hfill$
&&$ \hfill
0.01719  \pm 0.00004 \hfill$
&&$ \hfill 4.9\pm 0.1 \hfill$
&&$ \hfill -5.2\pm 0.3  \hfill$&\cr
height2.0pt& \omit&& \omit&& \omit&& \omit&& \omit&\cr
\fileth}}$$
\endinsert
\medskip
%
%
\subsection Parametric representation 

One then determines by the ODM method the  parameter $\rho$ and the function
$h(\theta)$, as explained in section 
\ssUQfixD. One obtains successive approximations in the form of polynomials of
increasing degree for $h(\theta)$. 
At leading order $h(\theta)$ is a polynomial of degree 5, whose coefficients are
given by the relations \eFHrelat{}:
$$h(\theta)=\theta[1+h_3(\rho)\theta^2+h_5(\rho)\theta^4]. \eqnd\ehtetv $$
For the range of admissible values for $F_5$ the coefficient $h_5$ of
$\theta^5$ given by eq.~\eFHrelat{b} has no real zero in $\rho$. It has a
minimum instead 
$$\rho^2=\rho_5^2={1\over12 F_5}(\gamma-2\beta) .\eqnd\erhomin$$
Substituting this value of $\rho$ into expression \ehtetv~one obtains the
first approximation for $h(\theta)$. At next order one looks for a minimum
$\rho_7$ of $|h_7(\rho)|$. One finds a polynomial either of degree 5 in
$\theta$, when  $h_7$ has a real zero, or of degree 7 when it has only a
minimum. It is not possible to go beyond $h_9(\rho)$ because already $F_9$ is too 
poorly determined.
\par
Note that one here has a simple test of the
relevance of the ODM method. Indeed, once $h(\theta)$ is determined,
assuming the values of the critical exponents $\gamma$ and $\beta$, one can
derive the corresponding function $F(z)$. It has an expansion to all orders in $z$. As a
result one obtains a prediction for the coefficients $F_{2l+1}$ which have not
yet been taken into account to determine $h(\theta)$. The relative difference
between the predicted values and the ones directly calculated gives an idea
about the accuracy of the ODM method. Indeed from the values $F_5=0.01711,
\gamma=1.2398, \beta=0.3256$, one obtains
$$F_7=4.83\times 10^{-4} ,\quad F_9=-3.2\times 10^{-5},\quad F_{11}=1.4\times 
10^{-7}\ \cdots \ . $$
We see that the value for $F_7$ is quite close to the central value
one finds by direct series summation, while the value for $F_9$ is within 
errors. This result gives us confidence in the method. It also shows that
the value of $F_9$ obtained by direct summation contains little new
information, it provides only a consistency check.
Therefore the simplest representation of the equation of state, consistent
with all data,  is given by
$$h(\theta)=\theta-0.76201(36)\;\theta^3+8.04(11)\times 10^{-3}\;\theta^5,
\eqnd\emainres $$
(errors on the last digits in parentheses) that is obtained 
from $\rho^2=2.8667\ $.
This expression of $h(\theta)$ has a zero at 
$\theta_0=1.154 $, which corresponds to the coexistence curve.
The coefficient of $\theta^7$ in eq.~\emainres~is smaller than $10^{-3}$.
Note that for the largest value of $\theta^2$ which corresponds to 
$\theta_0^2$, the $\theta^5$ term is still a small correction. 
Finally the corresponding values for the $\varepsilon$-expansion
are $h_3=-0.72$, $h_5=0.013$. These values are reasonably 
consistent, because a small change in $h_3$ can be cancelled to a large extent
by a correlated change in $\rho$. 
\medskip
The Widom scaling function $f(x)$, eq.~\ehscal, can then easily be obtained
numerically from $h(\theta)$ and compared with other determinations.
The main disagreement with other predictions comes from the region 
$x\rightarrow \infty$, i.e.~from the small magnetization region, where 
the predictions of the present method  should be specially reliable. 
\nref\rbervil{C. Bervillier, {\it Phys. Rev.} B34 (1986) 8141.}
\nref\rliu{A.J. Liu and M.E. Fisher, {\it Physica} A156 (1989) 35.}
\nref\raharony{A. Aharony and P.C. Hohenberg, {\it Phys. Rev.} B13 (1976)
3081; {\it Physica} 86-88B (1977) 611.}
\nref\rprivman{V. Privman, P.C. Hohenberg, A. Aharony, {\it Universal Critical
Point Amplitude Relations}, in Phase Transitions and Critical Phenomena
vol.~14, C.~Domb and J.L.~Lebowitz eds., (Academic Press 1991).}
\nref\rCasel{
M. Hasenbusch and K. Pinn, {\it J. Phys.} A31 (1998) 6157, cond-mat/9706003\semi
M. Caselle and M. Hasenbusch, {\it J. Phys.} A30 (1997) 4963, hep-lat/9701007\semi 
M. Caselle and M. Hasenbusch, {\it Nucl. Phys. Proc. Suppl.} 63 (1998) 613, 
hep-lat/9709089.}

\subsection Amplitude ratios

Various amplitude ratios can then be derived from
$h(\theta)$ and 
the values of the critical exponents determined earlier from longer series. They involve ratios of functions of $\theta$ at $\theta=0$ and at $\theta=\theta_0$ where $\theta_0$ is the zero
of $h(\theta)$ closest to the origin.\par
Tables \label{\ttcramp} and \label{\ttcrampb}  contain a comparison of several amplitude ratios as 
obtained from RG, lattice calculations (High temperature series and Monte-Carlo simulations) and experiments on binary mixtures, liquid--vapour, uniaxial magnetic systems.
\midinsert
\tablbl\ttcramp
$$ \vbox{\elevenpoint\offinterlineskip\tabskip=0pt\halign to \hsize
{& \vrule#\tabskip=0em plus1em & \strut\ # \ 
& \vrule#& \strut #
& \vrule#& \strut #  
& \vrule#& \strut #  
& \vrule#& \strut #  
&\vrule#\tabskip=0pt\cr
\noalign{\centerline{Table \ttcramp} \tableskip}
\noalign{ \centerline {\it Amplitude ratios.}
\tableskip} 
\fileth
height2.0pt& \omit&& \omit&& \omit&&\omit&& \omit&\cr
&$ \hfill $&&$ \hfill A^+/ A^-
\hfill$&&$ \hfill C^+/C^- \hfill$&&$ \hfill
R_c \hfill$&&$ \hfill R_\chi\hfill$&\cr
height2.0pt& \omit&& \omit&& \omit&& \omit&& \omit&\cr
\fileth
height2.0pt& \omit&& \omit&& \omit&& \omit&& \omit&\cr
&$ \hfill \varepsilon-{\rm exp.},\refs{\rbervil,\rNA}
 \hfill$&&$ \hfill 0.524\pm 0.010
\hfill$&&$ \hfill 
4.9 \hfill$&&$ \hfill  \hfill$&&$
\hfill 1.67 \hfill$&\cr
&$ \hfill \varepsilon-{\rm exp.}$,  \refs{\rGuiZJ,\rGuiZJii}  \hfill&
&$ \hfill 0.527\pm0.037 \hfill$&
&$ \hfill 4.73\pm 0.16 \hfill$&
&$ \hfill 0.0569\pm0.0035 \hfill$&
&$\hfill 1.648\pm0.036 \hfill$&\cr
height2.0pt& \omit&& \omit&& \omit&&
\omit&& \omit&\cr &$ \hfill d=3,~\rBBMN~\hfill$&&$ \hfill 0.541\pm
0.014 \hfill$&&$ \hfill  
4.77\pm 0.30\hfill$&&$ \hfill 0.0594\pm 0.001\hfill$&&$ \hfill 1.7
\hfill$&\cr 
height2.0pt& \omit&& \omit&& \omit&&
\omit&& \omit&\cr &$ \hfill d=3$,  \refs{\rGuiZJ,\rGuiZJii} \hfill
&&$ \hfill 0.537\pm0.019 \hfill$
&&$ \hfill  4.79\pm 0.10\hfill$
&&$ \hfill 0.0574\pm 0.0020\hfill$
&&$ \hfill 1.669\pm0.018\hfill$&\cr 
height2.0pt& \omit&& \omit&& \omit&& \omit&&\omit&\cr
&$ \hfill \hbox{HT series}~\rCPRV \hfill$&&$ \hfill 0.530\pm 0.003 \hfill$&&$ \hfill
4.77 \pm 0.02 
\hfill$&&$ \hfill 0.0564\pm0.0003 \hfill$&&$ \hfill 1.662\pm0.005
\hfill$&\cr 
height2.0pt& \omit&& \omit&& \omit&& \omit&&\omit&\cr
&$ \hfill \hbox{MC~\label{\rCasel}} \hfill$&&$ \hfill 0.560\pm0.010 \hfill$&&$ \hfill
4.75 \pm 0.03 
\hfill$&&$ \hfill  \hfill$&&$ \hfill \hfill$&\cr 
height2.0pt& \omit&& \omit&& \omit&& \omit&& \omit&\cr
&$ \hfill {\rm bin.~mix.} \hfill$&&$ \hfill 0.56\pm0.02
\hfill$&&$ \hfill 
4.3\pm 0.3 \hfill$&&$ \hfill 0.050\pm0.015  \hfill$&&$ \hfill  1.75\pm0.30
\hfill$&\cr  
height2.0pt& \omit&& \omit&& \omit&& \omit&& \omit&\cr
&$ \hfill {\rm liqu.-vap.} \hfill$&& \hfill 0.48--0.53 \hfill&& \hfill 
4.8{--}5.2 \hfill&&$ \hfill 0.047\pm0.010  \hfill$&&$ \hfill  1.69\pm0.14
\hfill$&\cr 
height2.0pt& \omit&& \omit&& \omit&& \omit&& \omit&\cr
&$ \hfill {\rm magn.~syst.} \hfill$&& \hfill 0.49{--}0.54 \hfill&&$ \hfill 
4.9\pm 0.5 \hfill$&&$ \hfill   \hfill$&&$ \hfill  
\hfill$&\cr 
height2.0pt& \omit&& \omit&& \omit&& \omit&& \omit&\cr
\fileth}}$$
\endinsert
\midinsert
\tablbl\ttcrampb
$$ \vbox{\elevenpoint\offinterlineskip\tabskip=0pt\halign to \hsize
{& \vrule#\tabskip=0em plus1em & \strut\ # \ 
& \vrule#& \strut #
& \vrule#& \strut #  
& \vrule#& \strut #  
&\vrule#\tabskip=0pt\cr
\noalign{\centerline{Table \ttcrampb} \tableskip}
\noalign{\centerline{\it Other amplitude ratios.}\tableskip}
\fileth
%
%
height2.0pt& \omit&& \omit&& \omit&&\omit &\cr
&\omit && $\hfill R_0 \hfill$&&$ \hfill R_3 \hfill$&&$ \hfill {C_4^+/
C_4^-}\hfill$&\cr  
%
%
height2.0pt& \omit  && \omit&& \omit&& \omit&\cr
\fileth
%
height2.0pt& \omit&& \omit&& \omit&& \omit&\cr
&$ \hfill {\rm HT\ series}~\refs{\rZLFish,\rCPRV}\hfill$&&$ \hfill  0.1275\pm0.0003\hfill$&&
$ \hfill 6.041\pm0.011 \hfill$&&$ \hfill -9.1\pm 0.2 \hfill$&\cr
height2.0pt& \omit&& \omit&& \omit&& \omit&\cr
& \hfill$ d=3$, \refs{\rGuiZJ,\rGuiZJii}\hfill&&$ \hfill  0.12584\pm0.00013\hfill$&&
$ \hfill 6.08\pm0.06 \hfill$&&$ \hfill -9.1\pm 0.6 \hfill$&\cr
& \omit&& \omit&& \omit&&  \omit&\cr
& \hfill $\varepsilon$-expansion, \refs{\rGuiZJ,\rGuiZJii} \hfill&&$ \hfill
0.127\pm0.002\hfill$&& 
$ \hfill 6.07\pm0.19 \hfill$&&$ \hfill -8.6\pm 1.5 \hfill$&\cr
& \omit&& \omit&& \omit&&  \omit&\cr
\fileth}}$$
\endinsert

The results from the $\varepsilon$-expansion  of \refs{\rbervil, \rNA} are obtained by direct Pad\'e summation of each corresponding series), 
while the results of \rGuiZJ~are obtained by first summing
$h(\theta)$ and then computing ratios, as explained in section \ssUQfixD. 
The results from the $d=3$ fixed dimension expansion of ref.~\rBBMN~ refer to direct summation up to $O(g^5)$ while the $d=3$ values of \rGuiZJii~again rely on the method explained in ref.~ \rGuiZJ. High temperature results are taken  from \refs{\rZLFish,\rCPRV} (see also
\refs{\rliu,\raharony}). Experimental data are  extracted from \rprivman, to which we
refer for more results and references. One notes the overall consistency of
the results obtained by different methods.
\section Concluding remarks

Within the framework of renormalized quantum field
theory and renormalization group, the presently available
series allow,  after proper summation, to determine precisely critical exponents for the $N$-vector model and the complete scaling
equation of state for 3D Ising-like ($N=1$) systems. In the latter example
additional technical tools, beyond Borel summation methods, are required in
which the parametric representation plays a central role. From the equation of
state new estimates of some amplitude ratios have been deduced which seem 
reasonably consistent with all other available data. \par 
Clearly a similar strategy could be applied to other quantities in a
magnetic field, in the scaling region. Note also that an extension of the
$\varepsilon$-expansion of the equation of 
state for $N=1$ to order $\varepsilon^4$ or even better $\varepsilon^5$,
that does not seem an unrealistic goal, would significantly improve the
$\varepsilon$-expansion estimates and would therefore be quite useful.
Finally the present approach could be extended to systems in the 
universality class of the $(\phi^2)^2_3$ field theory for $N\ne 1$, provided
expansions of the renormalized effective potential at high enough order can be
generated.   \par
More generally it must be emphasized that only a small number of universal quantities, as predicted by renormalization group, have yet been calculated. In addition to static infinite volume quantities, for which much work remains to be done, 
dynamic properties have not even be touched, finite size calculations would be useful for comparison with computer simulations.\par
Considering the increase in computer power more perturbative calculations will become feasible. What has been demonstrated here is that once the series are available, summation methods have been developed which lead to precise determinations.
\bigskip
{\bf Acknowledgments.} The author thanks R.~Guida for careful reading of the manuscript.
\nref\reqstreste{For other articles devoted to the equation of state see for example: J. Rudnick, W.  Lay, D. Jasnow, {\it Phys. Rev.} E58 (1998) 2902, cond-mat/9803026\semi
P. Butera, M. Comi, {\it Phys. Rev.} B58 (1998) 11552, hep-lat/9805025; 
{\it ibidem}\/ B60 (1999) 6749, hep-lat/9903010\semi
A. I.  Sokolov, E.V. Orlov, V.A. Ul'kov, S.S. Kashtanov, {\it Phys. Rev.} E60 (1999)  1344, 
hep-th/9810082\semi
J. Engels, T. Scheideler, {\it Nucl. Phys.} B539 (1999) 557\semi
S. Seide, C. Wetterich, {\it Nucl. Phys.} B562 (1999) 524, cond-mat/9806372\semi 
M. Strosser, S.A. Larin, V. Dohm, {\it Nucl. Phys.} B540 (1999) 654, cond-mat/9806103.}

\listrefs
\bye